\documentclass[hidelinks]{aa}

\usepackage{graphicx}

\usepackage{txfonts}

\usepackage{hyperref}
\usepackage{siunitx}
\usepackage{gensymb}

\usepackage{float}
\usepackage[caption = false]{subfig}

\usepackage{tablefootnote}

\newcommand{\citeg}[1]{\citep[e.g.][]{#1}}
\newcommand{\citsee}[1]{\citep[see][]{#1}}
\newcommand{\kms}[0]{{\,km\,s$^{-1}$}}
\newcommand{\msol}[0]{{\,M$_\sun$}}

\begin{document}

   \title{The physical association of planetary nebula NGC~2818 with open cluster NGC~2818A}

   \subtitle{}

   \author{{Vasiliki Fragkou}
          \inst{1}
          \fnmsep\thanks{Corresponding Author}, 
          {Roberto Vázquez}
          \inst{2},
          {Quentin A. Parker}
          \inst{3},
          {Denise R. Gonçalves}
          \inst{1},
          {Luis Lomelí-Núñez}
          \inst{1}
          }

   \institute{\\Observatório do Valongo, Universidade Federal do Rio de Janeiro, Ladeira do Pedro Antônio-43, Rio de Janeiro 20080-090, Brazil
              \email{vfragkou@ov.ufrj.br}
             \thanks{All inquires should be directed to the corresponding author}
         \and
            \\Instituto de Astronomía, Universidad Nacional Autónoma de México, 22800, Ensenada, B.C., Mexico
         \and
            \\Laboratory for Space Research, The University of Hong Kong, Cyberport 4, Pokfulam, Hong Kong, PRC
             }

   \date{}

  \abstract
   {Planetary nebulae (PNe) that are physical members of open star clusters (OCs) provide us with highly valuable data for stellar evolution studies. Unfortunately, they are extremely rare, with only three such instances previously confirmed in our Galaxy.}
   {In this study, we confirm the physical association of PN NGC~2818 with OC NGC~2818A, an association 
   long debated in literature studies. Hence, we add a fourth object to the sample of rare OC-PN pairs. The 
   physical properties of the PN can then be linked to those of its progenitor star.}
   {Using GHOST at Gemini high-resolution nebular spectra, we measured the PN systemic radial velocity to compare it with that of the putative host cluster. This was well determined from Gaia. We estimated the physical parameters of both the OC NGC~2818A and the PN using our data, together with estimates from previous studies and theoretical cluster isochrones and evolutionary tracks to show they are compatible.}
   {The highly precise, systemic radial velocity of the PN that was determined, is consistent to within the small errors with that of the cluster and its 
   1{\kms} associated velocity dispersion. This is a tight and primary requirement for cluster membership.  We have determined 
   other physical parameters of the PN and cluster, such 
   as age and distance. These also match within the errors. Taken together, these results present compelling evidence for 
   the physical association of the PN and cluster. The PN age was found to be around 11 kyr and the effective 
   temperature of its central star was estimated as 130 kK. The initial and final masses of the progenitor star were determined to 
   be $2.33\pm0.10${\msol} and $0.58\pm0.10${\msol}, respectively. We plotted the resulting initial-to-final data point on the latest initial-to-final-mass relation and also include the data points from our other three OC-PN associations known. This new data point agrees with the published trends from previous studies and further delineates the 'kink' found at relatively low initial masses.}
   {We show that all four OC-PN associations, identified thus far, share a number of common properties. These rare cases merit 
   detailed cluster-PN studies and work to further extend the sample.}

\authorrunning{Fragkou et al. 2025}
   \maketitle

\section{Introduction}
\label{sec:intro}
Planetary nebulae (PNe) are the ejected material associated with the final stages of evolution of low-to-intermediate stars (up to 8~M$_\sun$). Since these stars represent the vast majority of stars more massive than our Sun and dominate the chemical enrichment of our Galaxy, at least with respect to the light elements, their detailed study is of major importance for understanding how matter is distributed and recycled across the Galaxy. 

One major problem in PN studies is the lack of reliable data for the mass of their main sequence (MS) progenitors. If known, this would allow a direct determination of the link between their initial and final masses. While such an estimation is an unreliable task for general PNe, this is not the case for PNe
that are physically associated with star clusters. In these rare cases, the PN initial masses can be estimated with adequate precision from independent studies of the cluster turn-off mass via
colour-magnitude diagrams (CMDs) and theoretical cluster isochrone fitting.

Unfortunately, the PN phase is very brief, typically just a few tens of thousands of years \citep{2015ApJ...804L..25B}, though the oldest PN known also resides in an open cluster (OC) with a kinematic age of 78000$\pm$25000 years  but is an extreme, unique outlier \citep{2022ApJ...935L..35F}. OCs are relatively young, typically a few hundred million years old, and so only higher-mass progenitors can form PNe within such clusters \citeg{2019NatAs...3..851F}. PNe-open star cluster associations are consequently extremely rare. Indeed, up until now, only 
three PNe have very strong evidence to support their physical association with Galactic open star clusters \citep{2011MNRAS.413.1835P}. These associations are based on several key
%\citep{2011MNRAS.413.1835P,2019MNRAS.484.3078F,2019NatAs...3..851F,2022ApJ...935L..35F}. These associations are based on several key 
agreements. The most important is the tight agreement, within the small errors of the radial velocities of the PN and host cluster. This is because cluster members have a very  tight velocity spread to $\sim1${\kms} \citep{2000ASPC..198..517M}. Confirmation of membership is further supplemented by agreement of distance, reddening, proper motions, and even the PN physical size at the cluster distance being feasible so that taken together these measured parameters provide a compelling case for association.

It is valuable to establish more OC-PN associations and study them in detail because of their unique capability to allow for an independent determination of the properties of their progenitors.
Our new, high-resolution, spectral data for PN NGC~2818 provide high precision radial velocity measures across the PN to within 1\kms, which were then used to determine its systemic velocity
which, together with updated Gaia distances, reddening estimates, and other properties, demonstrates the likely physical association of PN~NGC~2818 with OC NGC~2818A. The current agreement of radial velocities to within 1 {\kms} between the PN and the cluster should be interpreted cautiously, due to the uncertainties inherent to the nebula’s systemic radial velocity determination. The tight match may be coincidental and fortuitous, given  the observed error margins, but it is worth noting that the derived systemic velocity aligns well with  previously reported values using different methods. We believe this adds to the evidence for cluster membership. This work should finally bring to an end the long debate about the status of this OC-PN association. 

The PN NGC~2818 was first detected by Herschel in the 19th century and has been studied in detail by numerous authors \citep[e.g.][]{1972MNRAS.158...47T, 1984ApJ...287..341D, 2012ApJ...751..116V, 
2024MNRAS.530.3327D}. \citet{1972MNRAS.158...47T} were the first to report a possible association of the PN with the OC given its line-of-sight angular proximity. This cluster, taking its official designation from the possibly associated PN, was also named NGC~2818A. 

\citet{1984ApJ...287..341D} conducted an early, comprehensive study of PN NGC~2818. The author found that it is a high-excitation PN with a reddening of $E(B-V)=0.23\pm0.02$, a radial velocity of 18.5$\pm$1.7\kms, an ionised mass larger than 0.6\msol, a radius of 1.1\,pc, an expansion velocity of $52\pm 3$\kms, a kinematic age of 22,000 years, and enhanced N and He  abundances, indicating it has type I PN chemistry \citep{1983IAUS..103..233P}. Assuming that the PN is a physical member of the cluster, the author estimated a PN progenitor mass of $2.2\pm0.3${\msol} \citet{2016MNRAS.455.1459F} estimated a nebular reddening of $E(B-V)=0.17\pm0.08$. We use this value throughout this analysis as it generally agrees with other estimates. \citet{2023ApJS..266...34G} also provide the most recent reddening estimate of $E(B-V)=0.186$ (with no error given) that agrees well with the value we have adopted and is in good agreement with other estimates.

In different studies, the PN distance estimates are varied between
2.50\,kpc \citep{1995A&A...293..541V}, 3.05\,kpc \citep{1995ApJS...98..659Z}, 1.79\,kpc 
\citep{2002ApJS..139..199P}, 2.00\,kpc \citep{2008ApJ...689..194S}, $2.50\pm0.50$\,kpc \citep{
2010ApJ...714.1096S}, $3.0\pm0.8$\,kpc \citep{2016MNRAS.455.1459F}, and 4.17\,kpc 
\citep{2021AJ....161..147B}. Agreement between PN and cluster distance estimates is a key determinant of cluster membership, so the greater than factor of two spread in PN distances among these 
works is an issue that needs to be resolved. The PN distance estimates are generally unreliable when based on indirect methods, and Gaia data for the PN central star (CSPN) where some of the aforementioned literature studies rely, are inconclusive due to its faintness (see below). We consider the best available statistical surface brightness radius (SB-r) relation PN distance of $3.0\pm0.8$\,kpc from \citet{2016MNRAS.455.1459F} as the most reliable PN distance determination due to the proven and relatively high accuracy ($\sim22$\%) for this well calibrated statistical distance scale.

\citet{1992A&AS...94..399C} estimate a nebular electron temperature of 15,200 K and a He/H abundance of 0.11 supporting the type I chemistry suggested by \citet{1984ApJ...287..341D}. \citet{2013MNRAS.431....2F} estimate an absolute nebular H$\alpha$ flux of $\log F({\rm H}\alpha)=-10.73\pm0.04\,$mW\,m$^{-2}$, while \citet{2023A&A...680A.104B} find an absolute nebular H$\beta$ flux of $\log\,F({\rm H}\beta)=-11.28\pm0.07\,{\rm mW\,m}^{-2}$, a N/H abundance of $8.4\pm0.2$ and a O/H abundance of $8.5\pm0.2$, also supporting the evidence that the PN is of Type~I chemistry. Moreover, recent ultraviolet (UV) studies  determine a nebular near-UV magnitude of $14.382\pm0.009$ and a far-UV magnitude of $14.387\pm0.015$ \citep{2023ApJS..266...34G}.

The identified blue CSPN is located at the projected geometric 
centre of the PN with ${\rm RA=09^h16^m01\fs7}$ and DEC\,$=-36\degr37\arcmin38\farcs8$, J2000 \citep{2011A&A...526A...6W}. Its Gaia DR3 magnitudes are $G=19.00\pm0.01$\,mag, $Bp=17.34\pm0.04$\,mag, and $Rp=17.11\pm0.05$\,mag, which, if converted to the Johnson-Cousins system\footnote{for the conversion we used the coefficients from the Gaia DR3 documentation, their Table 5.9.} are translated as $B=19.16\pm0.10$\,mag, $V=19.04\pm0.10$\,mag, and $R=18.94\pm0.08$\,mag. As described below, these faint magnitudes are problematic for Gaia estimates of distance.

The CSPN has a Gaia DR2 parallax of $0.27\pm0.32$\,mas (indicating a distance of 3.7~kpc) and a Gaia DR3 parallax of 
$0.03\pm0.21$\,mas (indicating a distance of 31.3~kpc). The Gaia DR3 parallax implies a very large distance that is not 
supported by the evidence and is very different than the one quoted in Gaia DR2. Moreover, very low, as in this case, and 
negative Gaia DR3 parallaxes are considered problematic and cannot be trusted. Gaia data \citep{2022A&A...667A.148G} have
large uncertainties for the very faint CSPN magnitudes of many PNe and indeed many are beyond Gaia limits\footnote{0.5\,mas parallax 
uncertainty for sources of $G=20$\,mag and 0.07\,mas parallax uncertainty for sources of $G=17$\,mag -see Gaia DR3 documentation.} ($G=19.00$\,mag, see above) so they are not reliable, including for this case, so we did not take these data into account in the following analysis. In early studies the CSPN effective temperature has been estimated to be 149\,kK \citep{2008ApJ...674..954B} and 160\,kK \citep{2016MNRAS.459..841M}, while its Zanstra \ion{H}{i} and \ion{He}{ii} temperatures were estimated to be 175\,kK and 215\,kK, respectively \citep{1988A&A...197..266G}.

Detailed morpho-kinematical PN studies with ShapeX \citep{ 2011ITVCG..17..454S} by \citet{2012ApJ...751..116V} and more recently by \citet{2024MNRAS.530.3327D} reveal that the nebula has a complex, non-uniform bipolar structure with multiple microstructures, bubbles, filaments, and cometary knots. Fabry-Perot data reveal a PN size of $120\times50$\,arcsec$^2$ as measured in the H$\alpha$ line and $160\times60$\,arcsec$^2$ in the [\ion{N}{ii}]$\lambda6584\AA$ line \citep{2024MNRAS.530.3327D}. The bipolar PN morphology, the same as found for all other confirmed OC-PN, is consistent with that associated with higher mass progenitors and those PNe emerging from common envelope binary evolution systems, including magnetic jet ejections misaligned to the axis of symmetry of the PN's main 
bipolar structure  \citep{2024MNRAS.530.3327D}. The inclination angle of the major axis of the PN, with respect to the line-of-sight, is estimated to be $60\degr$ by \citet{2012ApJ...751..116V} and $70\degr\pm3\degr$ by \citet{2024MNRAS.530.3327D}.

\citet{2012ApJ...751..116V} find a PN polar expansion velocity of 105{\kms} and an equatorial expansion velocity of 20\kms. Based on those, and a PN distance of 2.50\,kpc \citep{1995A&A...293..541V}, they estimate a PN kinematic age of $8400\pm3400$\,yr, much lower 
than the value found by \citet{1984ApJ...287..341D} \citet{2024MNRAS.530.3327D} find a somewhat higher expansion velocity of $120\pm20${\kms} for the bipolar nebular component and an expansion velocity of $70\pm20${\kms} for its equatorial component, much larger than previous estimates and also atypically large for PNe. Distinct expansion velocities are also estimated for the different mictrostructures of the PN. Based on their findings and adopting a PN distance of 4.17\,kpc \citep{2021AJ....161..147B}, \citet{2024MNRAS.530.3327D} derive a PN kinematic age of $9400\pm2900$\,yr. 

Such an association has been long debated across multiple studies. The PN radial velocity (a key parameter that could confirm its cluster membership when compared to that of the cluster) is estimated by different authors to range from $-16$ to +21\,{\kms} \citep{1967cgpn.book.....P,1972MNRAS.158...47T,1984ApJ...287..341D,
1988ApJ...334..862M,1998A&AS..132...13D}. Based on these values and the estimated cluster radial velocities from different 
studies, some authors claim that the PN is a cluster member \citep[e.g.][]{1972MNRAS.158...47T, 1984ApJ...287..341D}, while others claim that it is not \citep{2001A&A...375...30M,2007PASP..119.1349M,2008MNRAS.386..324B}. 
However, all these nebular radial velocity estimates are obtained from low-resolution spectral data and have large uncertainties. High-resolution spectral data and radial velocity precision of up to 1{\kms} are required for any reliable hypothesis regarding the status of this being a true PN-OC pair on this basis
\citep[see][]{2011MNRAS.413.1835P,1931ZA......2....1Z,2019NatAs...3..851F,2022ApJ...935L..35F}. \citet{2012ApJ...751..116V} estimate a nebular radial velocity of $26\pm2${\kms} from their high-resolution 
spectra, the most reliable result until now. Using the same data \citet{2024MNRAS.530.3327D} estimate a nebular radial velocity of $23\pm4${\kms}, with a spectral resolution of 11.5{\kms}, claiming that the PN is most probably a cluster member. 

In a recent study, \citet{2023ApJ...945...11R} estimate a cluster membership probability of the CSPN of only 11\%. Despite this, they
still claim that the nebula is possibly a cluster member since the CSPN proper motion is only $3\sigma$ from the cluster's mean and the space velocities of the two objects are very similar. 
Their claim is supported by the Gaia DR3 documentation, which cites that proper motion uncertainties are 0.07 mas\,yr$^{-1}$ at $G=17$ and 0.50 mas\,yr$^{-1}$ at $G=20$. Based on this assumption 
they estimate a CSPN progenitor mass of 2.10{\msol} and a CSPN final mass of 0.66\msol. Spectral energy distribution (SED) fitting allowed \citet{2023ApJ...945...11R} to estimate the CSPN effective temperature equal to $T_{\rm eff}=190\pm8$\,kK or even higher. The CSPN radius is estimated to be $0.026\pm0.002\,{\rm R}_{\sun}$.

In this work, we intend to show that the OC NGC~2818A does indeed host the PN NGC~2818 using newly obtained Gemini High-resolution Optical SpecTrograph (GHOST) at Gemini high-resolution spectra that yield a  precision in individual measures of $\simeq$~1~\kms.
In Section~{\ref{sec:OC}} we explore the properties of the OC NGC 2818A. In Section~{\ref{sec:obser}} we describe the observations, while in Section~{\ref{sec:res}} we present our results. Finally, in Section~{\ref{sec:disc}} we discuss our findings and in Section~{\ref{sec:conc}} we state our conclusions.

\section{The open cluster NGC~2818A}
\label{sec:OC}

\begin{figure}
      \centering
      \includegraphics[width=\hsize]{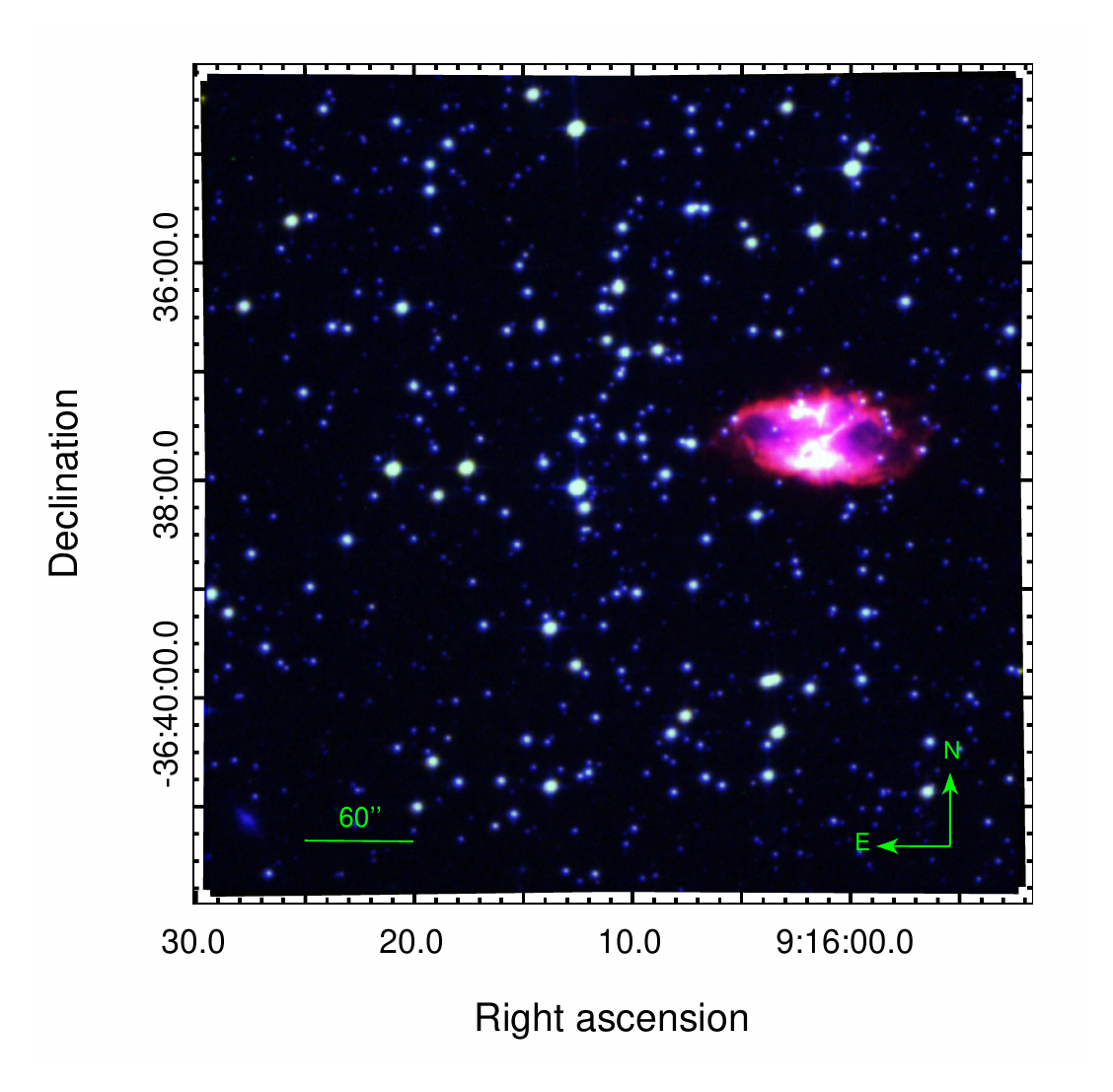}
      \caption{Open cluster NGC\,2818A. A $450\times450$\,arcsec$^2$ colour-composite RGB image of the cluster 
      NGC~2818A. Red is H$\alpha$, Green is short-red (SR) and Blue is broad-band blue as extracted from the UK Schmidt Telescope SuperCOSMOS H$\alpha$ Survey 
      \citep[SHS;][]{2005MNRAS.362..689P} through the Hong Kong/AAO/Strasbourg H$\alpha$ (HASH) planetary nebula database \citep{2016JPhCS.728c2008P}. The PN NGC\,2818, can be clearly seen on the west side of the cluster's apparent centre.
      }
      \label{Fig1}
   \end{figure}

The OC NGC~2818A is located along the line of sight of PN NGC~2818 (see Figure~\ref{Fig1}) and they share the same core identifier. The  $2400\pm300${\msol} \citep{2018MNRAS.480.3739B} cluster contains four blue stragglers and two yellow stragglers \citep[essentially evolved blue stragglers;][]{2023ApJ...945...11R}, and presents an extended MS turn off (TO) point, due to stellar rotation, and low levels of differential extinction \citep{2018MNRAS.480.3739B}. Its TO mass is 
estimated to be 2.1{\msol} \citep{1984ApJ...287..341D}. 

Multiple studies, listed below for convenience, have attempted to estimate the cluster physical parameters and these have been used here to estimate our reported mean cluster values from these literature determinations
\citep{1972MNRAS.158...47T, 1989AJ.....98.2146P, 1992A&AS...95..337T, 1995ApJS...98..659Z, 2000A&A...359..347D, 
2001A&A...375...30M, 2002NewA....7..553T, 
2002A&A...388..158L, 2006AJ....131.1559V, 
2010MNRAS.403.1491P, 2013A&A...558A..53K, 
2018A&A...618A..93C, 2018A&A...619A.155S, 2019ApJS..245...32L, 2019A&A...623A.108B, 2021ApJ...923..129J, 2021A&A...647A..19T, 
2021A&A...652A.102H, 2021MNRAS.504..356D, 2021MNRAS.502.4350S, 2022ApJS..259...19L, 2023ApJS..268...46L, 2023A&A...672A..29C, 
2024A&A...686A..42H, 2024AJ....167...12C}. Based on all these previously published values and estimating their error weighted average\footnote{for parameters with no assigned errors, we assume an error of 30~\%.}, we found a cluster systemic radial 
velocity of $22.0\pm2.6$\kms, cluster proper motions 
${\rm pmRA}=-4.4\pm0.3$\,mas\,yr$^{-1}$ and 
${\rm pmDEC}=4.5\pm0.3$\,mas\,yr$^{-1}$, a cluster mean parallax of $0.28\pm0.03$\,mas (this parallax, when converted, implies a distance of $3.6_{-0.4}^{+0.5}$\,kpc), a distance (as quoted in literature studies) of $3.5\pm0.2$\,kpc, a reddening $E(B-
V)=0.18\pm0.02$ (implying an extinction of $A_v=0.59\pm0.07$), an age of $1077\pm64$\,Myr, and an iron abundance of 
${\rm [Fe/H]}=-0.3\pm0.1$, which implies a metallicity of $Z=0.008\pm0.002$ for Z$_\sun=0.0152$ \citep{2012MNRAS.427..127B}. The \citet{1989ApJ...345..245C} extinction law with Rv=3.1 was used where relevant.

In recent work, \citet{2023A&A...675A..68V} conduct a comprehensive cluster study using Gaia DR2 data \citep{2016A&A...595A...1G}, to identify the stars that have a high probability to be cluster members. Following their findings and using only stars that have a probability $\ge90$\% to be physical members of NGC\,2818A (244 stars) and Gaia DR3 
data \citep{2022A&A...667A.148G}, we found a systemic cluster radial velocity of $22.0\pm1.3${\kms} (the quoted error was derived by standard error propagation of the Gaia measurement errors) with cluster member's radial velocities having a standard deviation of $\sigma=5.4$\kms, a mean cluster parallax of $0.303\pm0.003$\,mas (implying a cluster distance of $3.30\pm0.03$\,kpc), and cluster proper motions ${\rm pmRA}=-4.421\pm0.003$\,mas\,yr$^{-1}$ and 
${\rm pmDEC}=4.548\pm0.003$\,mas\,yr$^{-1}$. The derived parameters are in excellent agreement with the error-weighted mean of those reported by previous authors and since their estimation relies on reliable Gaia data and highly probable cluster members, we consider them as a better representation of the cluster true values and use these for the subsequent analysis. 

\begin{figure}[!]
   \resizebox{\hsize}{!}
            {\includegraphics{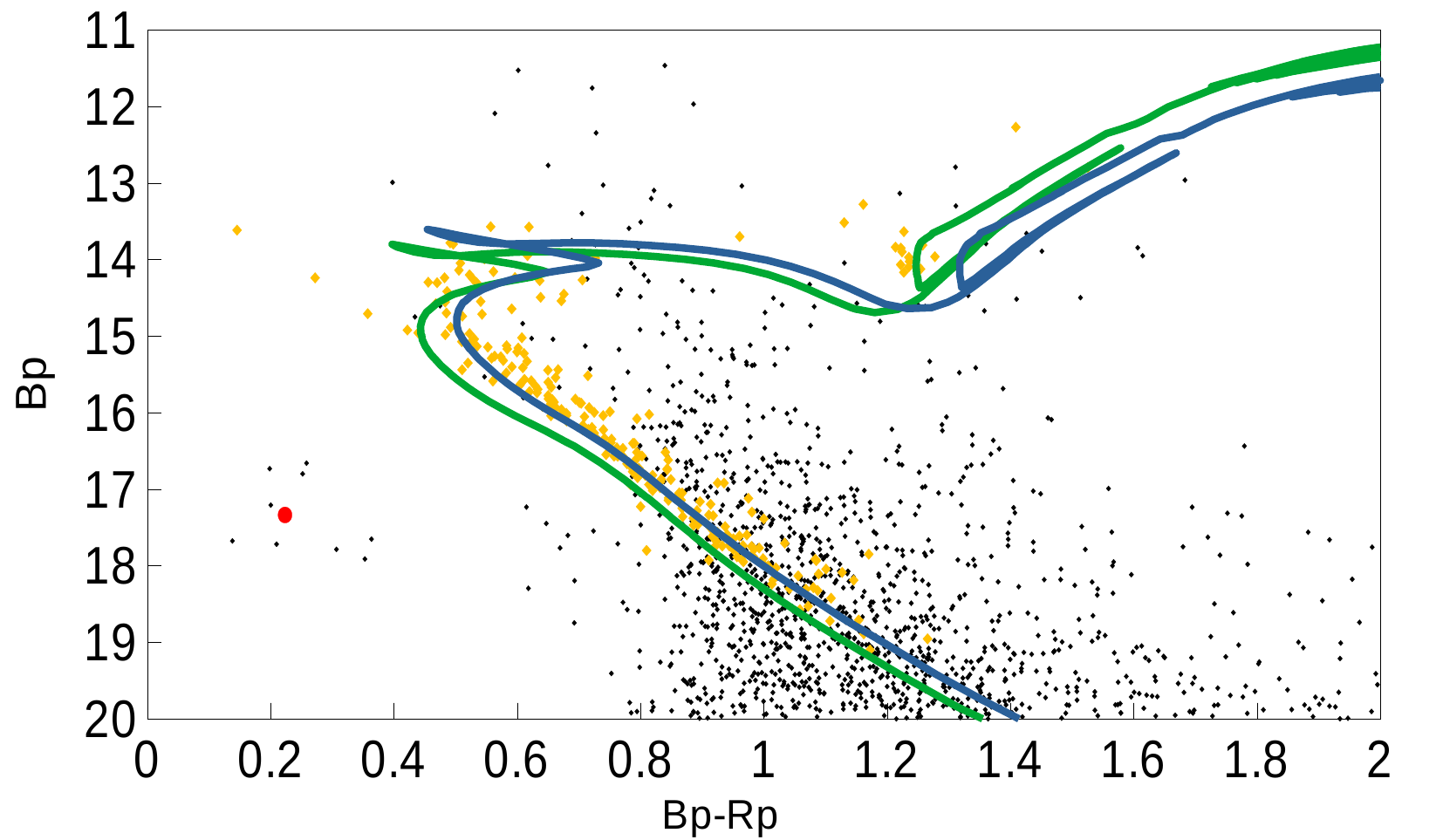}}
      \caption{NGC~2818A CMD. Padova theoretical isochrones overplotted to our cluster Gaia DR3 CMD for stars within 225\,arcsec from the cluster apparent centre. The orange
      points represent the stars with a cluster membership probability of $\ge90\%$ \citep{2023A&A...675A..68V}. The green isochrone reflects the error weighted average of literature values and Gaia DR3 distance (age$=1077$\,Myr, 
      $E(B-V)=0.18$, distance= 3.30\,kpc, iron abundance ${\rm [Fe/H]}=-0.29$), while the blue isochrone is the model that best fits to our high probability membership data (age= 955\,Myr, $E(B-V)=0.18$, distance= 3.30\,kpc, iron abundance ${\rm [Fe/H]}=-0.024$). The red dot shows the position of the identified CSPN, located well within the white dwarf region of the cluster as expected.}
         \label{Fig2}
   \end{figure}

We used, uncorrected for extinction, Gaia DR3 data for all stars within the cluster's tidal radius of 225\,arcsec from the 
cluster apparent centre, to construct a Gaia ($B_p-R_p$ vs $B_p$) cluster colour-magnitude diagram (CMD, Figure~\ref{Fig2}). High probability \citep[$\ge$ 90\%;][]{2023A&A...675A..68V} cluster members were overplotted in orange dots. Subsequently, we fitted in green a theoretical Padova isochrone \citep{2012MNRAS.427..127B,2013MNRAS.434..488M, 
2014MNRAS.444.2525C,2015MNRAS.452.1068C,2017ApJ...835...77M,2019MNRAS.485.5666P,2020MNRAS.498.3283P}, based on the cluster parameters derived above. We note that the isochrone does not fit very well to the data. Keeping the, highly reliable, Gaia OC distance ($3.30\pm0.03$\,kpc) and the consistent literature reddening $E(B-V)=0.18\pm0.02$, and using the \citet{1989ApJ...345..245C} extinction law with Rv=3.1, we performed additional fits of multiple isochrones varying the cluster age and metallicity. 
Applying a $\chi^{2}$ minimisation to our $\ge90\%$ membership probability data and the different models we found that the isochrone that best fits our data is for an age of $955\pm70$~Myr and an iron abundance of ${\rm [Fe/H]}=-0.024\pm0.015$ (metallicity $Z=0.0144\pm0.0005$ for Z$_\sun=0.0152$). The quoted errors reflect the best fit model if accounting for all Gaia DR3 data, within 225\,arcsec from the cluster's apparent centre, and the distance and reddening uncertainties.

To specify, the age and metallicity of the cluster were determined using a $\chi^{2}$ minimisation. The $\chi^{2}$ is defined as

    \begin{equation}
        \chi^{2} = \frac{1}{\nu} \sum_{i=1}^{n}\left(\frac{{\rm colour}_{{\rm DR3}} - {\rm colour}_{{\rm mod}}}{\sigma_{{\rm DR3}}}\right)^{2},
        \label{equ:chi2}
    \end{equation}
%\nointent
\noindent 
where ${\rm colour_{DR3}}$ and ${\rm colour_{mod}}$ are the Gaia DR3 and Padova models colours respectively, ${\rm \sigma_{DR3}}$ is the uncertainty of ${\rm colour_{DR3}}$ and ${\rm \nu=n-1}$ are the degrees of freedom. The value of ${\rm n}$ is the total number of stars in Gaia DR3 that have a cluster membership probability of $\ge90\%$. We performed Monte Carlo simulations adding Gaussian noise along with their associated uncertainties into the Gaia DR3 data. A Gaussian distribution of ages for the best ﬁt models, where the mean of the distribution represents the age of the cluster and its uncertainty, is given by the standard deviation  of the age distribution. The metallicity of the cluster corresponds to the metallicity of the model that provides the best ﬁt. We use these cluster values for the remainder of our study. From the best fitted isochrone we estimated a cluster TO mass of 1.88 M$_\sun$.

\section{Observations}
\label{sec:obser}

Confirming an OC-PN physical association requires the estimation 
of their systemic radial velocities with very high precision ($\leq$ 1~{\kms}) due to the very low 1~{\kms} cluster velocity dispersions seen in OCs. The OC radial velocity is well determined with Gaia, but there are only few similar studies about the PN. 
\citet{2012ApJ...751..116V} analysed high-resolution PN spectra and find a PN systemic radial velocity close to that of the cluster, but not precise enough to confirm cluster membership. We acquired new, highly precise, high-resolution GHOST at Gemini 8.2~m spectra (Program: GS-2023B-FT212) for 11 different pointings across the PN to provide a more robust estimate for this study. 

Uncertainties for individual pointings reflect the spread in radial velocities calculated from different PN emission lines. Aperture positions were carefully selected to be representative across this quite extended and evolved PN and so for reflecting the gas movements through the full extent of the nebula permitting an estimation of the systemic velocity through the average. The PN has a complex structure and different velocity components. Thus, the gas is moving in multiple directions with different speeds. For calculating the overall systemic PN velocity we had to take into account the movement of these nebular components at some level, given we have been using small integral field unit (IFU)-fibre samples and not a large IFU that can sample the full PN to provide an integrated value. Radial velocities computed from such symmetrically selected apertures counterbalance the gas's internal movement allowing for a more precise determination of the systemic radial velocity. The observed spread of radial velocities at each pointing across the PN is because they reflect the internal gas movement and not just the systemic radial velocity, which for most compact PNe is seen in the integrated spectra determined values. Recall that typical PN expansion velocities are around 25-35 {\kms} about the systemic value calculated -whereas this PN has a much higher expansion velocity. Here, the issue is resolved by selecting geometrically symmetric apertures throughout the PN that should largely 'neutralise' the internal gas movement. The aperture that is most representative of the PN's systemic radial velocity is pointing f that is centred on the PN's apparent centre and central star, and thus does not reflect internal gas movements in different directions and may be considered a reasonable proxy for the systemic velocity and agrees well with the cluster value. Unfortunately, due to the faintness of the CSPN, it would be very difficult to obtain deep CSPN spectral data of high resolution in order to compute an accurate CSPN radial velocity even with today's best ground based facilities. For this we have to wait for the next class of telescopes and surveys.

The GHOST spectrograph was configured in the dual-IFU standard-resolution mode, which can observe two targets simultaneously by using two small and independent $1.2\times1.2$\,arcsec$^2$ IFUs. GHOST's range (383-1000~nm), is ideal for PN spectral studies since it contains their strongest optical emission lines. It consists of two separate detectors, one for the blue and one for the red arm, which collect light simultaneously. Its resolution in the standard resolution mode is normally $R=56000$, more than enough for our purpose of measuring the PN's radial velocity at each selected point to individual precisions down to 1~{\kms}, but for our specific observations, this precision was a little lower than this (see below).

\begin{figure}
   \centering
   \includegraphics[width=\hsize]{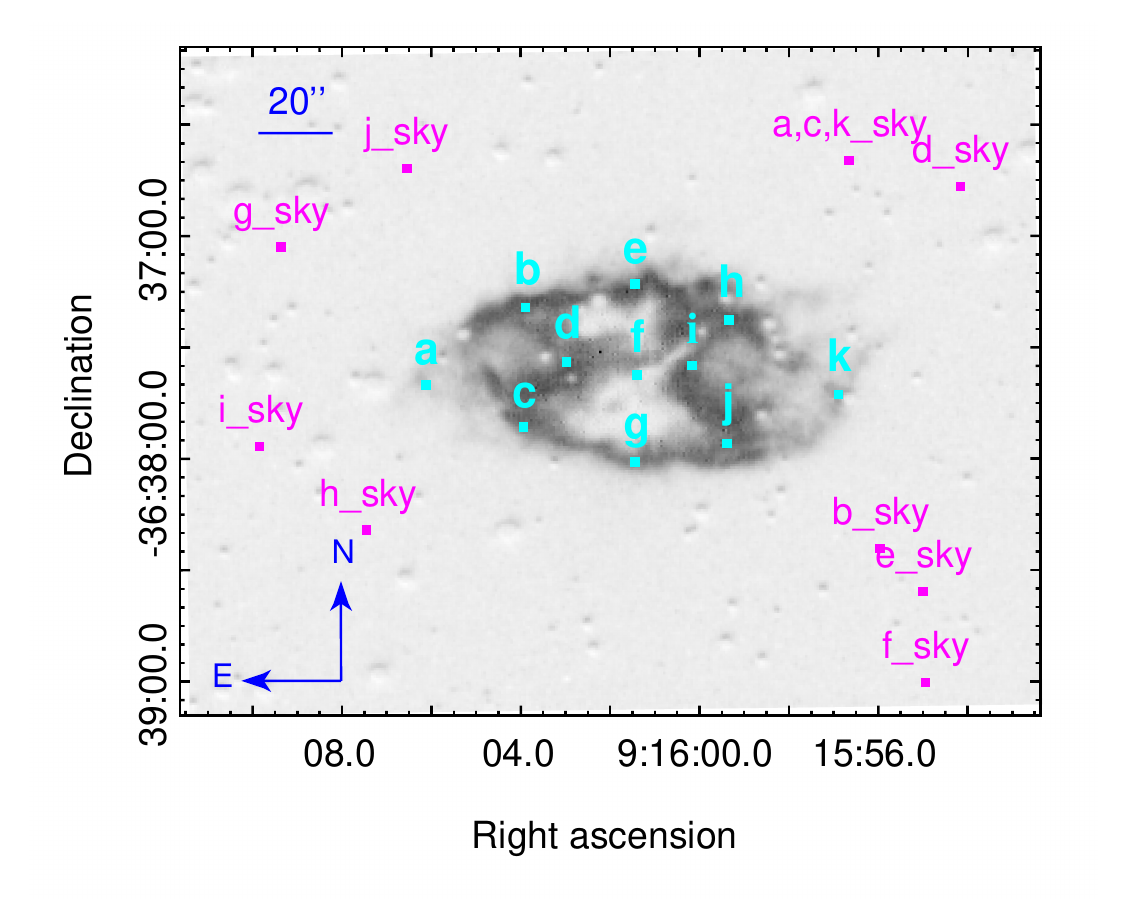}
      \caption{Planetary nebula NGC\,2818. A continuum subtracted H$\alpha$-SR SHS PN image indicating the position of the 11 GHOST $1.2\times1.2$\,arcsec$^2$ IFU PN (a-f) and their respective sky pointings. The pointing sizes represent the size of the GHOST IFUs.
              }
         \label{Fig3}
   \end{figure}

With ten 300-second exposures, we obtained spectra for 11 nebular pointings (labelled a-k), plus ten reference pointings in the adjacent sky for sky-subtraction (Figure~\ref{Fig3}). PN pointings a and k have been observed simultaneously and for their sky-subtraction the respective sky of the c pointing was used. For
our pointings we intentionally carefully selected areas free from stars to avoid any stellar contamination in our data. Pointing f is centred on the CSPN, but it is too faint to be detected in our relatively short exposures, even with the 8~m Gemini telescope.

We used a binning of $1\times1$, a slow readout, an up-looking port and silver mirror coating (the only available option at the time of the observations). The spatial binning of one of the GHOST data is a noticeable improvement in relation to the \citet{2012ApJ...751..116V} data that have a spatial binning of two. The pixel size in the spatial direction is 0.4~arcsec. The airmass of our targets was better than 1.5 for all exposures.

The data were acquired on December 09 and 10, 2023 in a shared-risk GHOST run. This was affected by some problems with the focus (3\% offset from the nominal value). The effects of this issue are minimal for observations in the dual target mode, where one of the IFUs is pointed on the sky, as in this work for most of our pointings. This is not the case for our exposure for pointings a and k, but since we used a separate sky IFU for their sky subtraction the effect is the same. A few adjustments were made in the reduction process for these out-of-focus data. These included i) setting the full-width-half-maximum (FWHM) of the Gaussian smoothing kernel to 6 pixels when reducing the flats, to account for the spatial smoothing of the slit profile caused by the poor focus and ii) extracting the spectra with uniform weighting. These steps improved the quality of the final data products despite  a 10-20\% reduction in the resolution ($R\approx40000$), still more than enough for our purposes (see Section~\ref{sec:res}). The reduced data are free of artefacts down to the 1\% level and for our analysis we do not use lines in the blue end of the spectrograph, where the dispersed orders are close enough to each other on the detector to cause possible cross-talk.

For the reduction of our data we used a combination of the DRAGONS 3.20 pipeline, as optimised for GHOST data acquired during the December 2023 shared risk run (see above), and standard Image Reduction and Analysis Facility (IRAF) techniques. For the wavelength calibration we used Thorium-Argon arc wavelength calibration lamps obtained before and after our science exposures. As our PN data have no continuum it was not possible for DRAGONS to estimate the seeing from the slit viewing camera images and fit an appropriate model to the individual fibre fluxes that is required for the extraction of the spectra. For this reason, we constructed a synthetic slit viewing camera image, as a Moffat profile of FWHM=5\arcsec to mimic uniform brightness, and used that instead. Since GHOST is primarily designed for point-sources (where the spatial profile is the same in all wavelengths) and because of the way that DRAGONS detects cosmic rays (based on 
deviations from the standard spatial profile), extended emission line sources may present some problems with cosmic-ray flagging when using a fairly uniform profile as in our case (if an emission line exists in only some of the fibres, it may get flagged as a cosmic ray and be removed from the final spectrum). Inspecting pixels flagged as cosmic rays in our data did not identify any significant issue. Barycentric correction has also been applied to our data and flux calibration was turned off, since it is not required for our science case.

The DRAGONS's final products (i.e. the 1D spectral data of all of our PN and sky pointings, part of which are presented in Figure~\ref{Fig4}) have been transformed to the non-standard format that IRAF uses for log-linear wavelength axes. For the sky subtraction, we used the IRAF, IMAGES package task IMARITH, their respective sky pointings for pointings b-j and the respective sky 
pointing of pointing c for pointings a and k.

\section{Results}
\label{sec:res}

Using the IRAF task {\sc splot}, we applied Gaussian fits to the stronger nebular emission lines, the [\ion{N}{ii}]$\lambda6548,6584\AA$ doublet and the H$\alpha$ line,  determining their peak wavelength, their Doppler shifts and hence radial velocities. We did the same for the [\ion{S}{ii}]$\lambda6716,6731\AA$ doublet for estimating the nebular electron density. Pointing f is centred on the location of the CSPN but, as expected, it is too faint for any relevant absorption lines to be detected in our short exposure spectral data. 

\begin{figure*}[!]
\subfloat[pointing $a$]{\includegraphics[width = 2.5in]{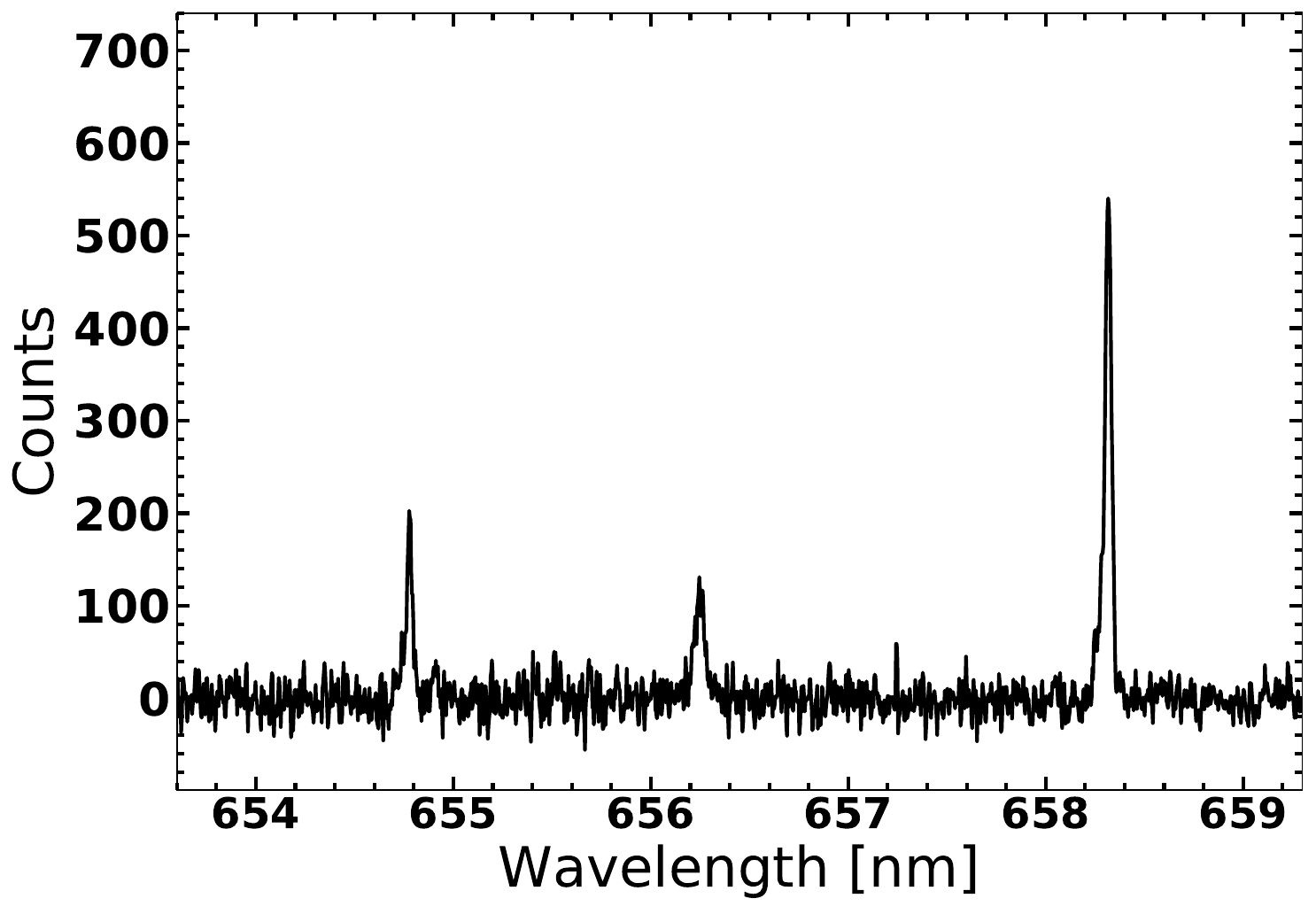}}
\subfloat[pointing $b$]{\includegraphics[width = 2.5in]{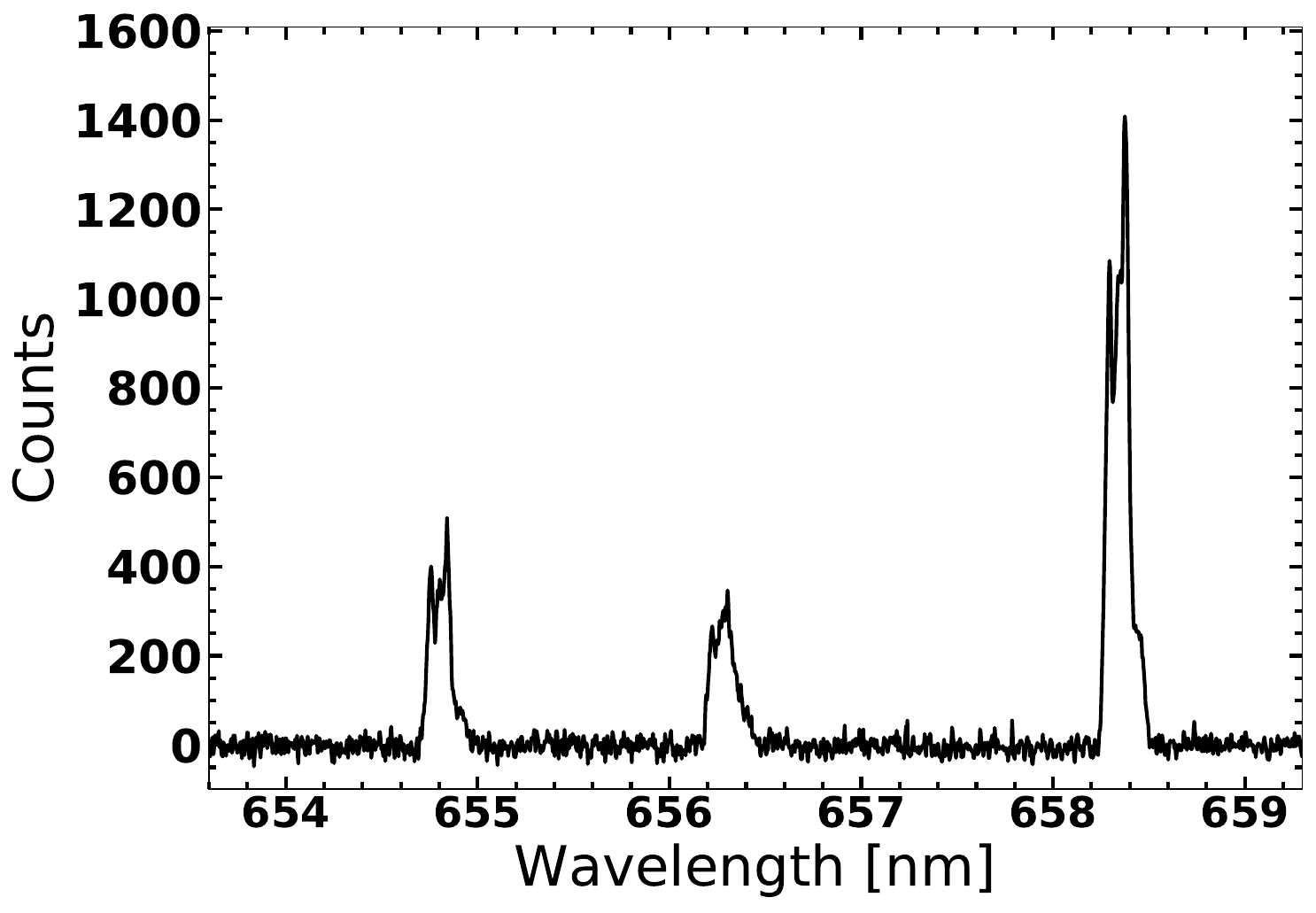}}
\subfloat[pointing $c$]{\includegraphics[width = 2.5in]{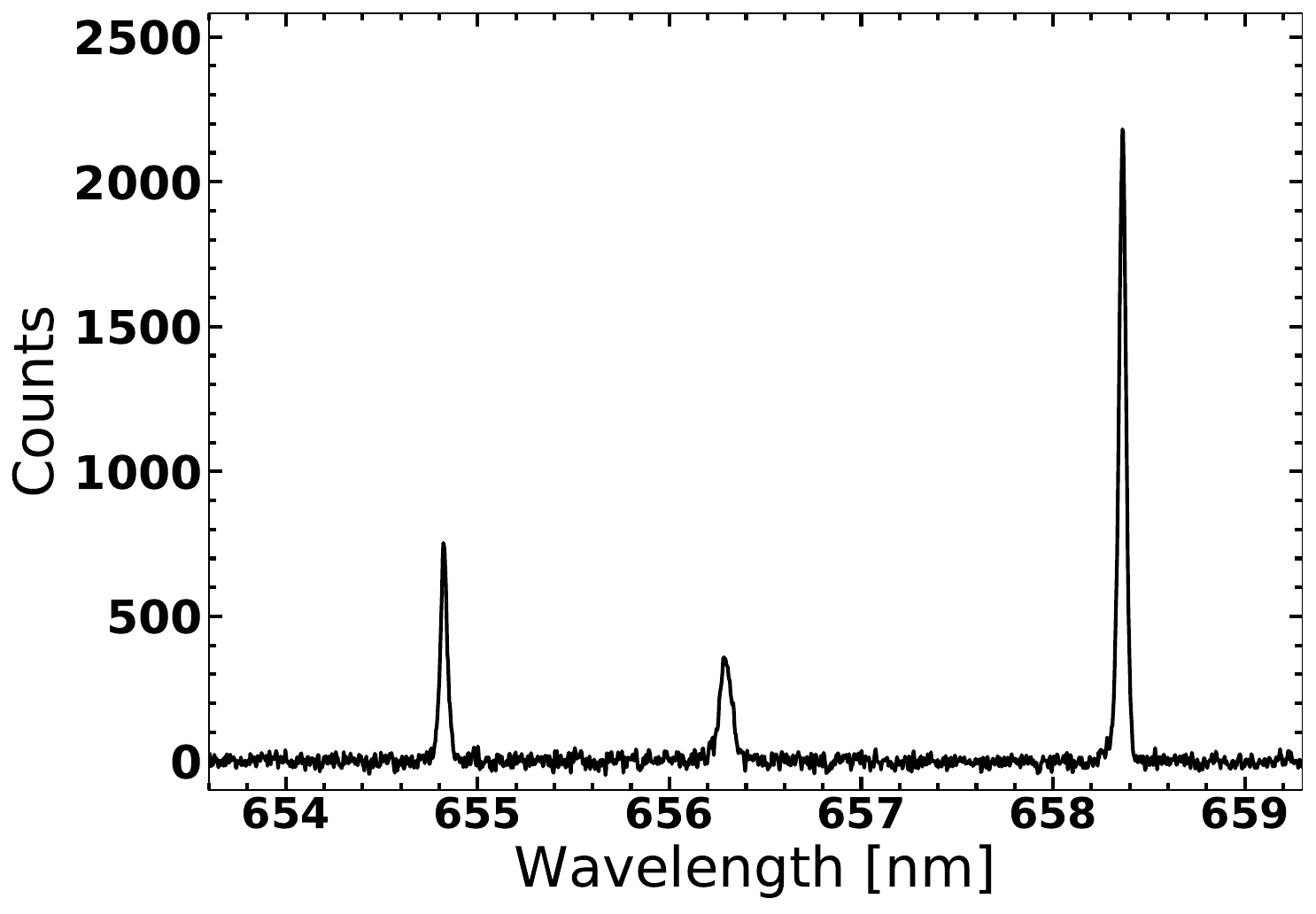}}\\
\subfloat[pointing $d$]{\includegraphics[width = 2.5in]{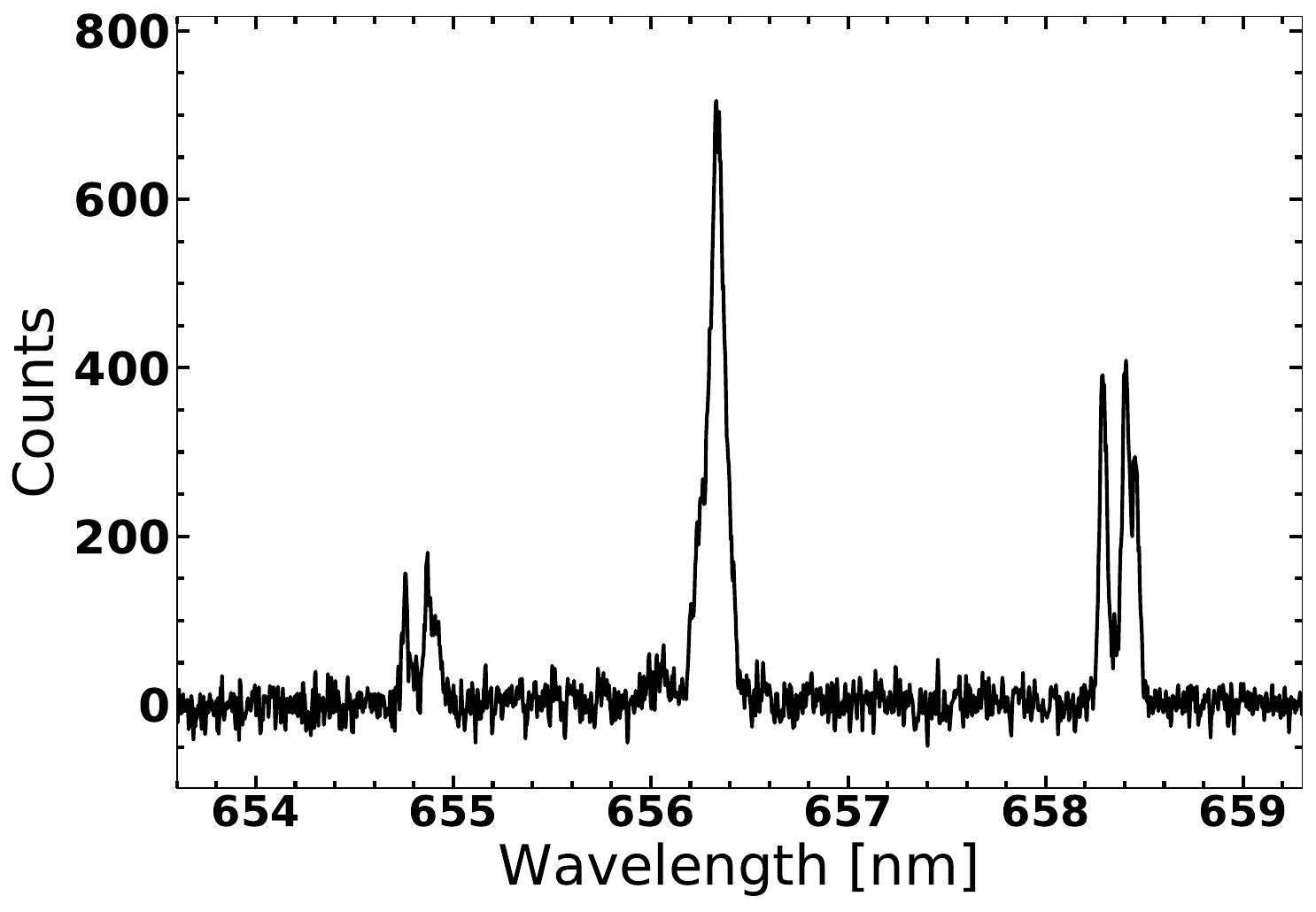}} 
\subfloat[pointing $e$]{\includegraphics[width = 2.5in]{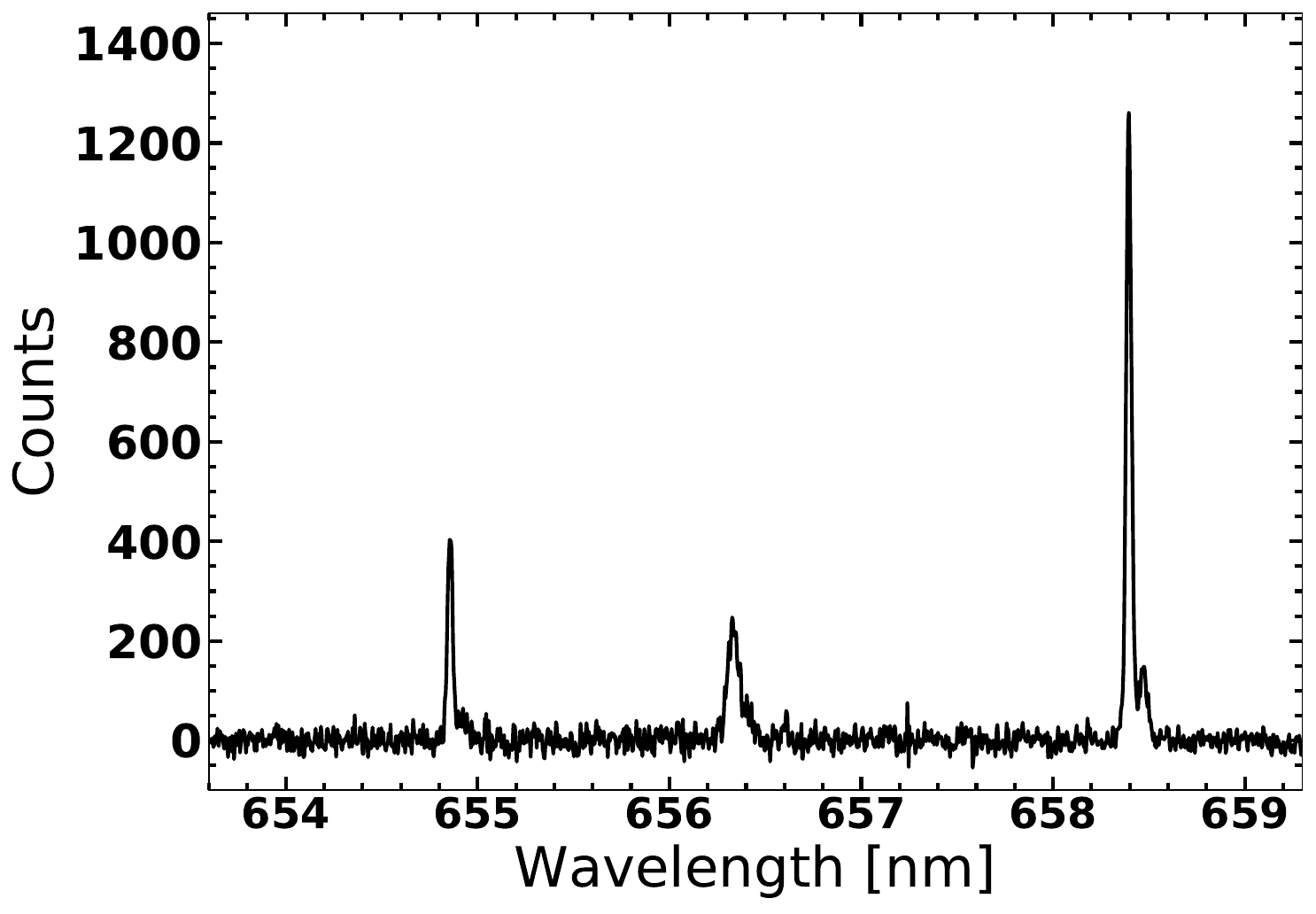}} 
\subfloat[pointing $f$]{\includegraphics[width = 2.5in]{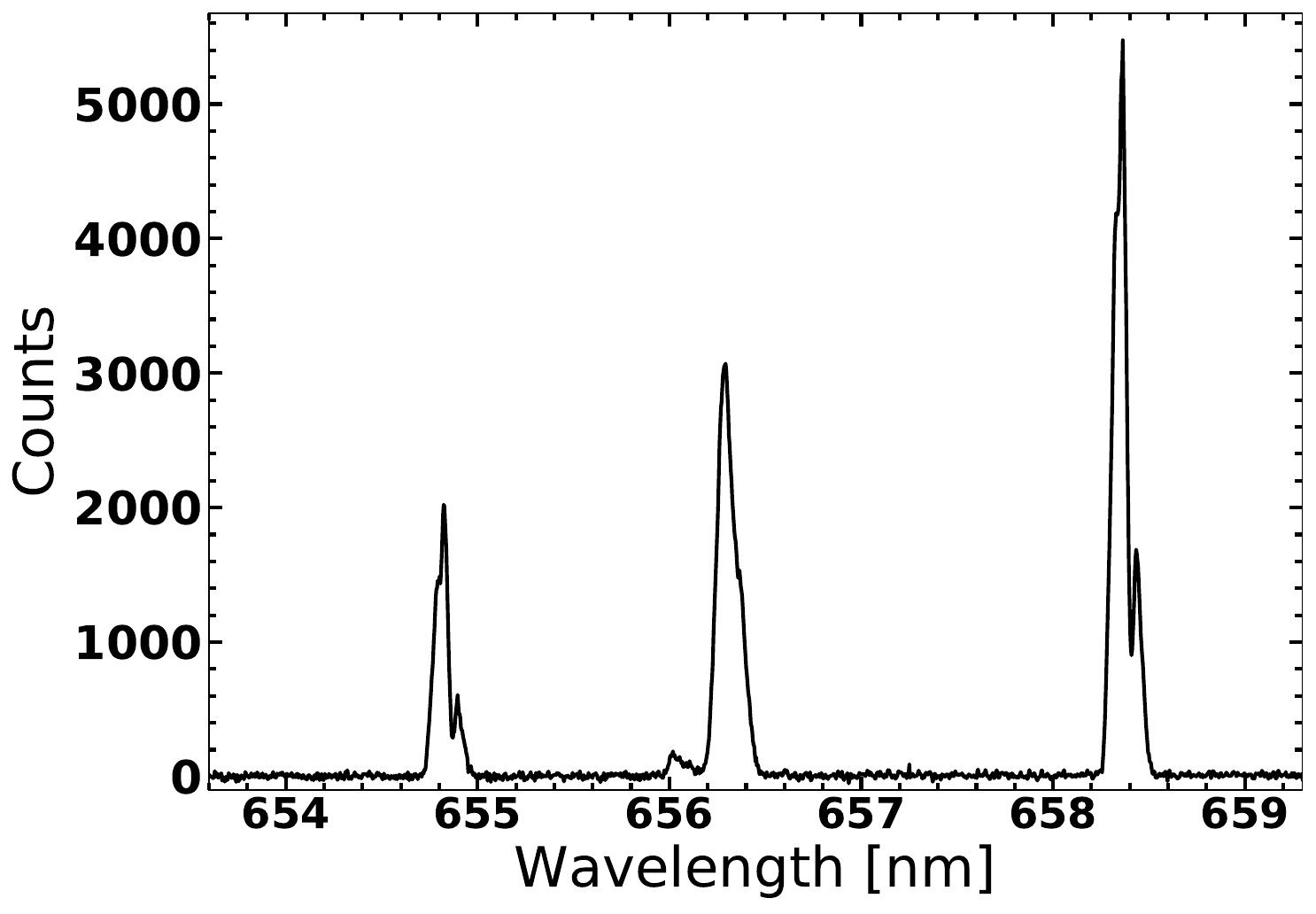}}\\
\subfloat[pointing $g$]{\includegraphics[width = 2.5in]{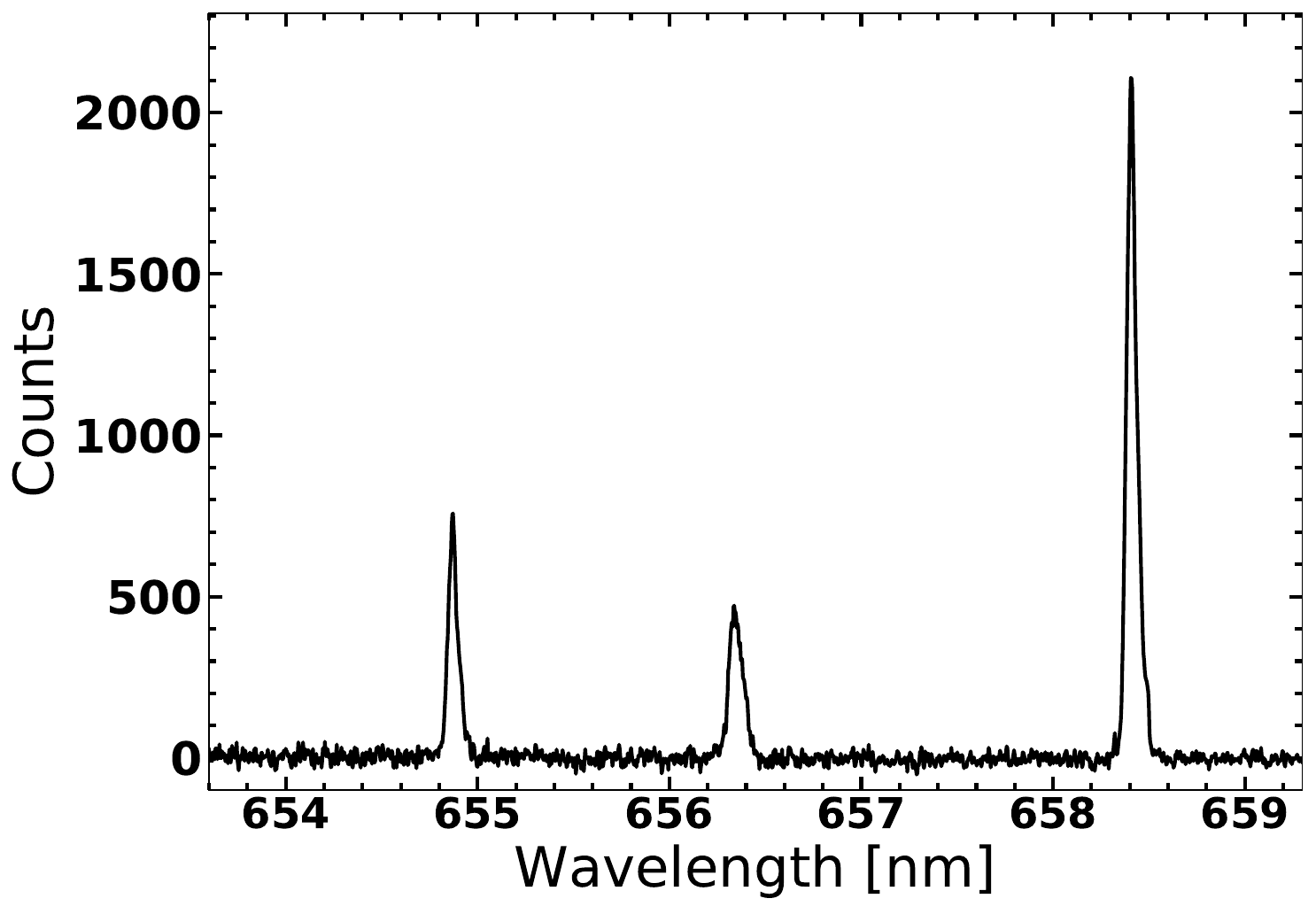}}
\subfloat[pointing $h$]{\includegraphics[width = 2.5in]{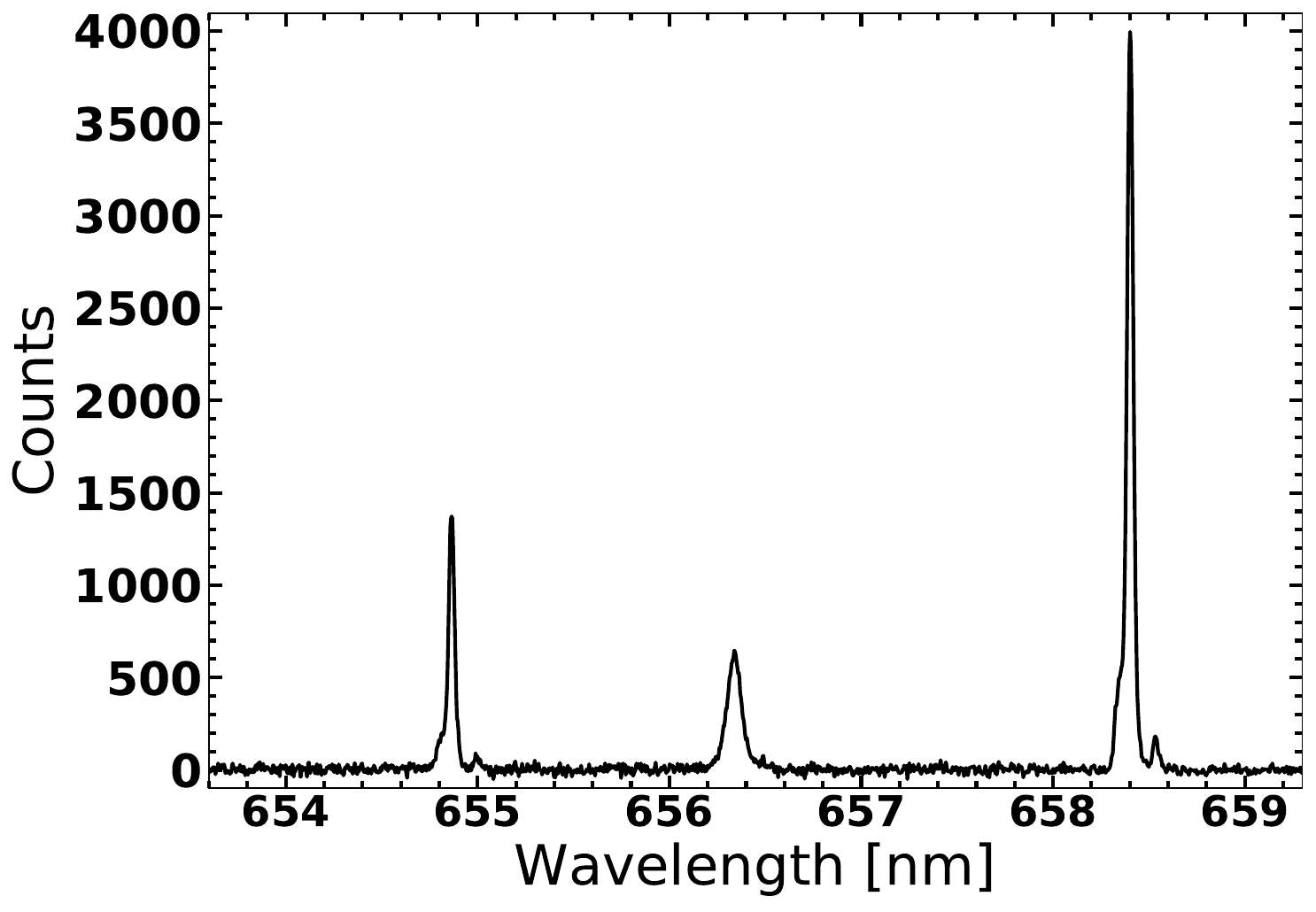}} 
\subfloat[pointing $i$]{\includegraphics[width = 2.5in]{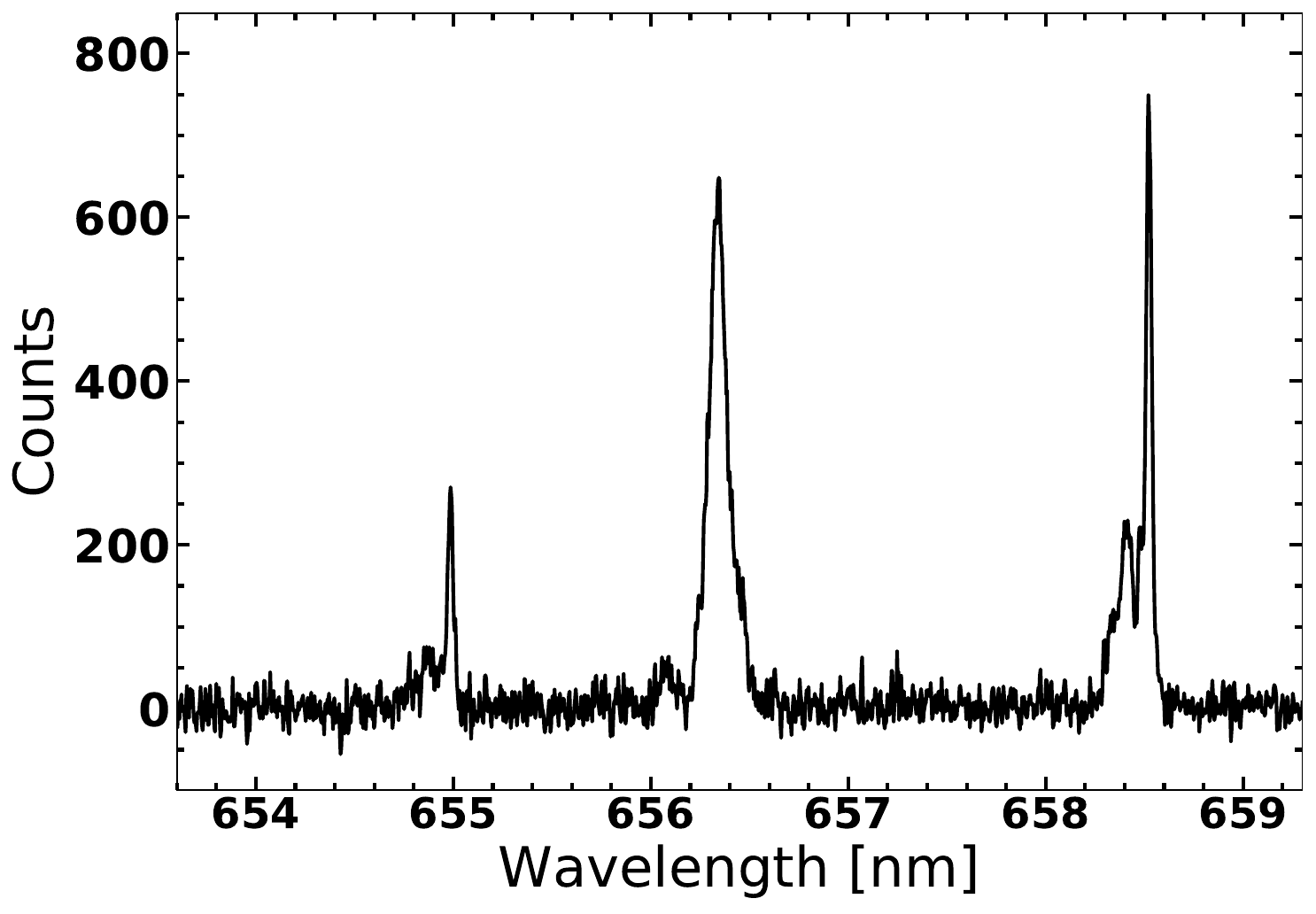}} \\
\centering
\hspace{1in}\subfloat[pointing $j$]{\includegraphics[width = 2.5in]{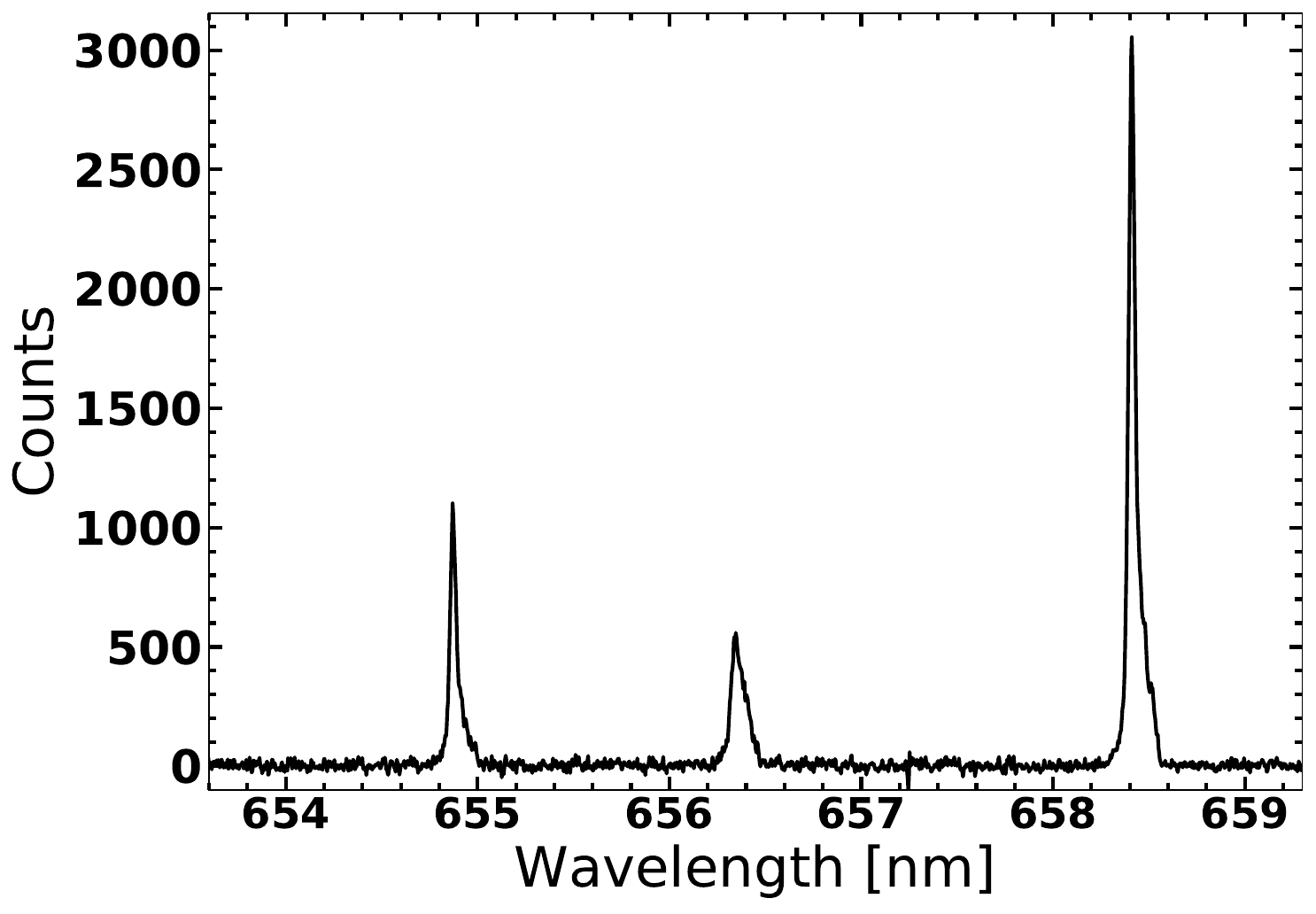}} 
\hspace{1.2in}
\subfloat[pointing $k$]{\includegraphics[width = 2.5in]{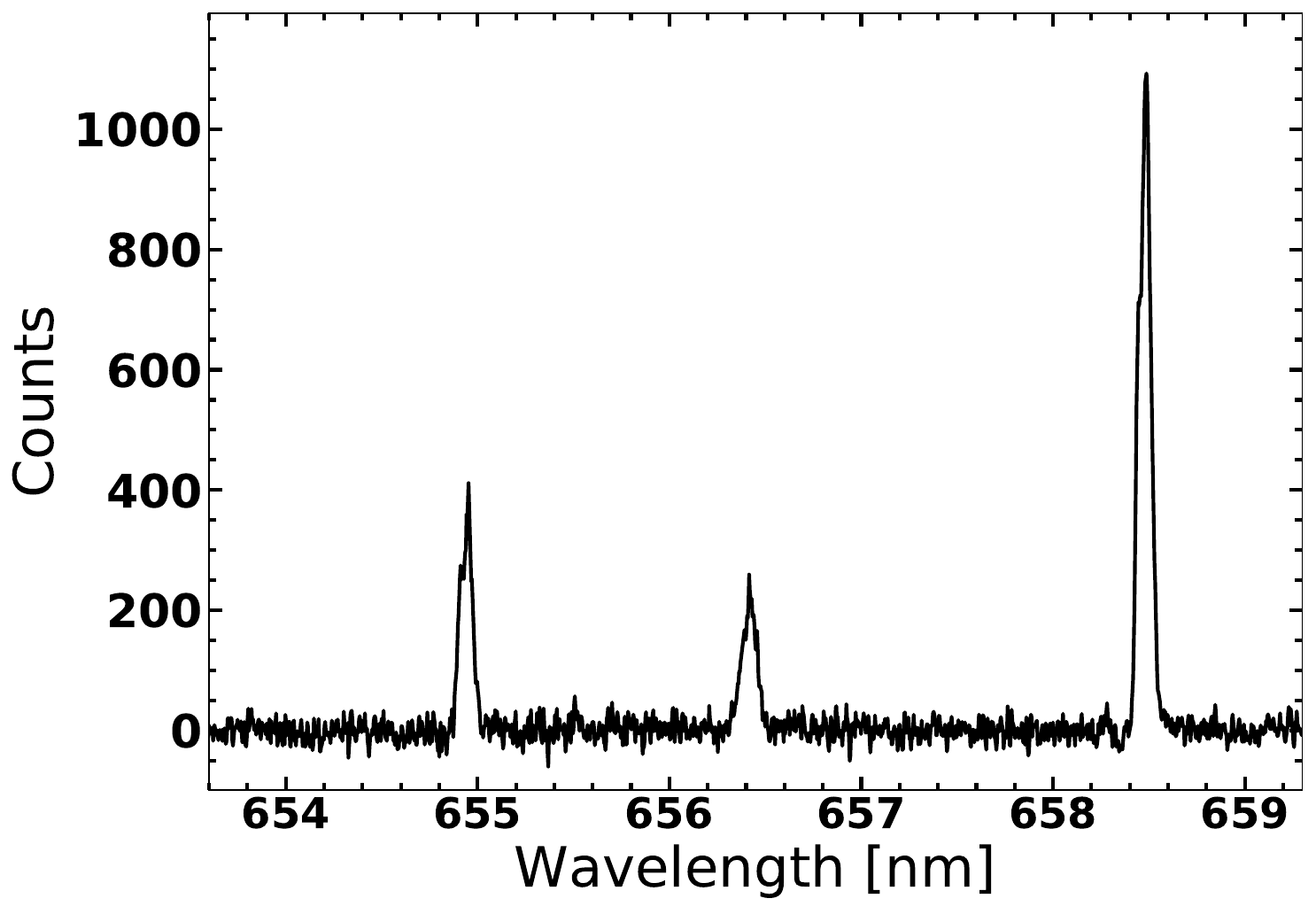}}
\caption{GHOST spectral data. The 1D spectra showing the H$\alpha$ and [\ion{N}{ii}]$\lambda6548,6584\AA$ doublet for each pointing a-k are provided. In the spectra of pointings b, d, e, f, and i we can see a prominent split of the [\ion{N}{ii}]$\lambda6548,6584\AA$ lines and in the pointings b, d, and f a less clear split of the H$\alpha$ line.} 
\label{Fig4}
\end{figure*}

In the spectra of pointings b, d, e, f, and i, we found that the 
[\ion{N}{ii}]$\lambda6548,6584\AA$ lines are split indicating significant nebula expansion at these sampled locations. The same 
applies for the H$\alpha$ line for pointings b, d and f, but the split is less significant (see Figure~\ref{Fig4}). The high-excitation structure found by \citet{2012ApJ...751..116V} in the central part of the nebula is also reflected in our data with the detection of the [\ion{He}{ii}]$\lambda4686\AA$ and [\ion{He}{ii}]$\lambda6560\AA$ lines in the spectra of pointings d, f and i.

From the $\lambda6300\AA$ sky line of our sky exposures, which are not contaminated from nebular emission, we can measure the resolution of the instrument and the precision of the measured radial velocities. Since this line is a sky line, its radial velocity should be zero. Any deviation from this value indicates the instrumental resolution. Using all our sky pointings, we found that our data have a mean resolution of 0.08 with $\sigma=0.03$ {\kms}. It is clear that the instrumentation provides more than enough precision to better than 1 {\kms} in any one measurement, confirming a much better radial velocity precision than the one required for our study. The issue then is how we can use our observations to determine the best systemic velocity estimate for the PN as a whole given our distributed pointings across this resolved source and the internal velocity structure and expansion velocity differences depending on the axis used and 
corruption for inclination angle.

The radial velocity for each pointing was measured from the Doppler shifts of the brightest nebula emission lines: 
the H$\alpha$ and the [\ion{N}{ii}]$\lambda6548,6584\AA$ doublet. High-resolution spectra allow the peaks of the individual components of the split spectral lines to be resolved and thus, measured separately. This give us a more precise determination of the radial velocity. In lower resolution data the peak of the lines would be biased towards the brightest component giving a false estimate \citep{2012ApJ...751..116V}. Three of our pointings (b, d, and f) present splits in all three H$\alpha$ and [\ion{N}{ii}] lines, while pointings e and i, present splits only in the case of the [\ion{N}{ii}] lines. The spectra of the rest of our pointings exhibit no split lines. In the case of pointings where 
all the lines present a split, we calculated the average radial velocity from both components and then estimated the mean radial velocity from all lines. In the case of pointings where only the 
[\ion{N}{ii}] lines are split, we estimated their radial velocity from their average, omitting the H$\alpha$ line, since its single peak is biased to its brightest component inducing large errors. In the case of pointings where no lines present a split, we estimated their radial velocities, as the average of the Doppler shifts of each H$\alpha$ and [\ion{N}{ii}] line, based on their single peaks. Table~\ref{Tab1} shows our radial velocity estimates for each pointing. 

By taking the average of the estimated radial velocities for each pointing, we found the systemic nebular radial velocity to be equal to $23.7\pm5.5$\kms, matching, within the errors, the cluster's well determined radial velocity. This provides key support to PN cluster membership. No matter which emission lines are chosen for measuring the radial velocity of each pointing, the final result of the systemic nebular radial velocity remains between 22 to 24{\kms} but with different induced measurement errors. If we only take into account the more accurate split emission line results 
(completely omitting pointings with only single peaks), the systemic nebular radial velocity was found to be $24.5\pm4.1${\kms}. Radial velocity uncertainties for individual pointings, reflect the radial velocities calculated from different PN emission lines. Moreover, and as described in Section~{\ref{sec:obser}}, the spread of radial velocities, as calculated from different pointings (see Table~{\ref{Tab1}}), reflects the nebular internal gas movement. Our data are not flux calibrated, plus the blue arm H$\gamma$ and H$\delta$ lines are too faint to be detected so it is not possible to estimate the nebular reddening from our blue arm spectra using H$\beta$ with these lines. However, recent independent estimates of the PN reddening \citep[e.g.][]{2016MNRAS.455.1459F,2023ApJS..266...34G}, agree very well with the determined cluster reddening of $E(B-V)=0.18$ adding further evidence to the case of a true PN-OC 
association. The reddening estimates, in conjunction to the 
\citet{2016MNRAS.455.1459F} surface-brightness radius relation PN statistical distance of 3\,kpc also agrees to the Gaia DR3 cluster distance, providing further support of PN cluster membership. 

\begin{table*}
   \caption{Estimated radial velocity, expansion velocity, [\ion{N}{ii}]/H$\alpha$ ratio, [\ion{S}{ii}] $I(6731)/I(6716)$ ratio, and $N_e$ for each pointing}             
   \label{Tab1}      
   \centering                          
   \begin{tabular}{c c c c c c}       
   \hline \hline
      \vspace{0.1cm}                 
      pointing & $v_{\rm r}$ $\pm$ $\sigma$ ({\kms}) & $v_{\rm exp}$
      \tablefootmark{\rm (1)} ({\kms}) & [\ion{N}{ii}]/H$\alpha$ & $I(6731)/I(6716)$ & $N_e$ \tablefootmark{\rm (2)} (cm$^{-3}$)  \\  
      \hline                        
      a & -13.8 $\pm$ 1.2 & 75 & 3.33 & 0.21 $\pm$ 0.04 & < 100\\ 
      b & -8.8 $\pm$ 2.8 & 100& 4.77& 0.64 $\pm$ 0.05 & < 100 \\
      c & 7.5 $\pm$ 1.7 & 85& 4.39 & 0.91 $\pm$ 0.04 & 497\\
      d & 5.4 $\pm$ 2.1 & 98 & 0.91 & 0.75 $\pm$ 0.08 & < 100\\
      e & 38.4 $\pm$ 0.3 & 63 & 3.35 &0.80 $\pm$ 0.03 & 230\\ 
      f\tablefootmark{\rm (3)} & 23.3 $\pm$ 2.2  & 93 & 1.83& 0.90 $\pm$ 0.01 & 471\\
      g & 30.5 $\pm$ 0.9 & 93 & 4.18 & 0.59 $\pm$ 0.03 & < 100 \\
      h & 26.5 $\pm$ 1.1 & 98 &3.66 &0.72 $\pm$ 0.02 & < 100\\
      i & 59.1 $\pm$ 0.4 & 92 & 1.09& 0.83 $\pm$ 0.13 & 297\\ 
      j & 30.4 $\pm$ 2.0 & 106 & 3.78 & 0.77 $\pm$ 0.02 & 165\\
      k & 62.1 $\pm$ 1.3 & 92 & 5.44 & 0.78$\pm$ 0.04 & 185\\ 
      \hline                                 
  \end{tabular}
\tablefoot{(1) Refers to the polar expansion velocity. (2) Assuming $T_e$= 11500 K. (3) centred on the CSPN and so also a reasonable proxy for the systemic PN velocity.}
\end{table*}

The nebular expansion velocity (see Table~\ref{Tab1}), was measured in each pointing from both the FWHM  of the 
H$\alpha$ line (taking into account for the thermal broadening using the $\lambda6300\AA$ night sky line) and the 
split of the [\ion{N}{ii}] lines where present. The measured values were then de-projected based on the average (65\degr) of the nebular inclination angles found by \citet{2012ApJ...751..116V} and \citet{2024MNRAS.530.3327D}. The 
mean nebular polar expansion velocity was found to be equal to 91{\kms} with $\sigma= 12${\kms}, close but somewhat lower than the values found by \citet{2012ApJ...751..116V} and \citet{2024MNRAS.530.3327D} (105~{\kms} and 120 $\pm$ 20~{\kms} respectively) by ShapeX morpho-kinematical modelling, while its equatorial expansion velocity was found to be 42 {\kms}, presenting a standard deviation of $\sigma$= 6 {\kms}. The higher expansion velocity estimates found in previous studies, are possibly a result of their employed models. The expansion velocity is interestingly atypically high for PNe.

\begin{figure}
   \centering
   \includegraphics[width=\hsize]{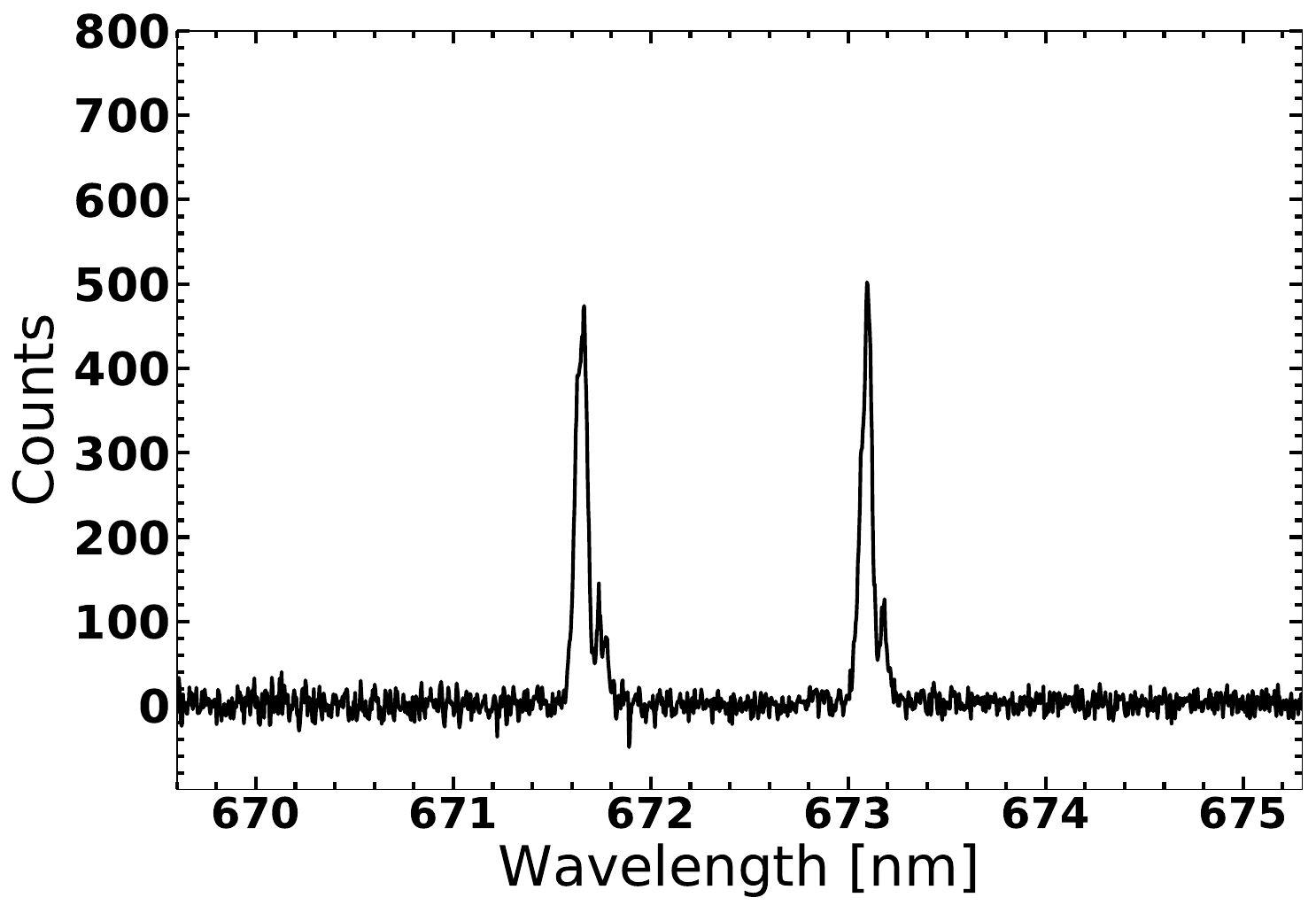}
      \caption{1D GHOST spectrum of pointing f (centred on the CSPN) showing the [\ion{S}{ii}]$\lambda~6716,6731\AA$ doublet used to estimate $N_e$. Note the mild line splitting as seen in other species for other pointings.
              }
         \label{Fig5}
   \end{figure}

The average [\ion{N}{ii}]/H$\alpha$ line ratio was found to be 3.34 - which is typical for Type-I PNe \citep[see][]{1994MNRAS.271..257K}. Such PNe are also mostly bipolar, as in the present case, and indeed all PNe so far confirmed to lie in OCs are bipolar. The average [\ion{S}{ii}]$\lambda6731/6716\AA$ ratio is used for determining electron densities (see Figure~\ref{Fig5}). For pointings b, d and i, the [\ion{S}{ii}] lines present a split, and  each component has been measured separately and then added. We obtain [\ion{S}{ii}]$\lambda6731/6716\AA$= $0.72\pm0.19$ (see Table~\ref{Tab1}). 

Considering that [\ion{N}{ii}] as well as [\ion{S}{ii}] are low-ionisation lines, whose ionisation potentials are close if compared, for instance, with that of [\ion{O}{iii}], $T_e[\ion{N}{ii}]$ is the appropriate $T_e$ to derive $N_e[\ion{S}{ii}]$. The $T_e[\ion{N}{ii}]=11500\pm800$\;K, found by \citet{1984ApJ...287..341D} and the above [\ion{S}{ii}]$\lambda6731/6716\AA$ are then adopted to obtain $N_e$. By using the Temden task in PyNeb \citep{2015A&A...573A..42L}, we estimated a mean nebular electron density of $N_e<100\,cm^{-3}$ though it is clear from Table~\ref{Tab1} that for several pointings, such as c and f in particular, that the $N_e$ is well determined at the level of a few hundred electrons revealing real point dependent variation. The derived electron density for some pointings is an upper limit because in these cases the [\ion{S}{ii}] line ratio is close to the low density limit, as shown by \citet{2006agna.book.....O}, and therefore the [\ion{S}{ii}] lines are not sensitive enough for calculating with precision a $N_e$.

Even though in the spectra of pointings f and i the [\ion{O}{iii}]$\lambda4363\AA$ line is detected, which would allow the direct derivation of $T_e[\ion{O}{iii}]$, our spectra are neither flux-calibrated nor corrected for interstellar extinction, therefore only  diagnostic emission-line ratios based on close enough lines, such as [\ion{S}{ii}]$\lambda$6731/6716$~\AA$, can be safely used for deriving the nebular physical parameters. The large spread of electron densities in different pointings (see Table~\ref{Tab1}) reflects the low signal-to-noise of the [\ion{S}{ii}] lines in some of our pointings and real density variations of different nebular components. The mean [\ion{S}{ii}]$\lambda6731/6716\AA$ ratio was used for the determination of the mean nebular electron density as the optimal representation of the true nebular physical conditions for the calculation of the nebular ionised mass below. Density variations in individual nebular components are not expected to affect these general conditions especially since many components exceed or approach the low density limit. For these reasons, we used the mean nebular electron density $N_{\rm e}<100\,{\rm cm}^{-3}$ where relevant.

\section{Analysis and discussion}
\label{sec:disc}

We adopt a PN apparent size of $140\times55$\,arcsec$^2$ as the mean of its major and minor axis in H$\alpha$ and 
[\ion{N}{ii}]$\lambda6584\AA$ imaging \citep{2024MNRAS.530.3327D}. Using the adopted Gaia cluster distance of $3.30\pm0.03$\,kpc we estimated a PN physical size of $2.24(\pm0.02)\times0.88(\pm0.01)$\,pc.

From the estimated nebular physical sizes (major and minor axes) and measured polar and equatorial expansion velocities we can determine the nebular kinematic age. We calculated separately the nebular age from its major axis and polar expansion velocity and from its minor axis and equatorial expansion velocity and then found their mean. Assuming that the expansion velocity has not changed with time, we estimated a PN kinematic age $\tau_{\rm 
k}=11000\pm2000$\,yr, slightly higher, but agreeing within the errors with age estimates from previous studies 
\citep{2012ApJ...751..116V,2024MNRAS.530.3327D} that assume different PN distances. Using the  effective recombination coefficient of H$\beta$ from \citet{2006agna.book.....O}, the absolute nebular H$\beta$ flux of $\log F({\rm H}\beta)= -11.28\pm0.07$\,mW\,m$^{-2}$ \citep{2023A&A...680A.104B}, the adopted cluster distance and the measured electron density, and following \citet{2015ApJ...803...99C}, we estimated an approximation of the nebular ionised mass larger than 0.47~M$_\sun$. The calculated ionised mass is large 
compared to typical PNe that can partially be explained by the nebular complex structure, since for determining the ionised mass using the available data we had to assume spherical geometry and homogeneous structure. The PN NGC~2818 is complex in both geometry and different structures. Thus, the determination of the ionised mass here is an approximation based on the available data. A more precise determination of the ionised mass would require detailed three-dimensional data encompassing the nebula's full spatial extent. Previous morpho-kinematic studies of this PN using 3D modelling tools, such as ShapeX \citep{2011ITVCG..17..454S}, show the significant role of non-uniform structures and varying expansion velocities in shaping its observed characteristics  \citep{2012ApJ...751..116V, 2024MNRAS.530.3327D}. While these models provide a deeper understanding of the nebula’s morphology and dynamics, extending them to derive ionised mass with high precision would  require integrating additional observational datasets, such as integral field spectroscopy, which is  beyond the scope of this study.

The PN has a relatively high and variable H$\alpha$ surface brightness, especially at certain locations, as evident in Figure~\ref{Fig3}. This is especially noteworthy given its large, physical size. Some reasons that can explain this are its relatively young age and relatively high CSPN luminosity, complex structure with multiple components and underlying mechanisms, and possible binary CSPN evolution (see Section {\ref{sec:intro}). The relatively high CSPN luminosity and temperature (given the presence of He~II emission in the PN) imply that the CSPN has recently entered the white dwarf (WD) cooling track, still having considerably potential of ionising the nebular gas, resulting in a relatively high surface brightness of the PN. Common envelope binary evolution would imply that the ejected material was mostly distributed along the axis of symmetry of the binary orbit \citep{2024arXiv241106831J} and thus, the gas concentrated mostly in specific geometrical areas. This could explain both the nebular high surface brightness present at certain components and its bipolar shape.

Knowing the CSPN V magnitude, as converted from Gaia DR3 photometry (see Section {\ref{sec:intro}), and using the 
adopted distance and PN reddening, we found the CSPN absolute magnitude to be $M_v=5.9\pm0.3$. By plotting the CSPN 
absolute magnitude and the derived nebular kinematic age along with the corresponding (i.e. for $Z=0.01$) \citet{2016A&A...588A..25M} evolutionary tracks, we estimated a CSPN final mass of $0.58\pm0.01$\msol. The quoted error reflects the uncertainties in the derivation of the CSPN absolute magnitude and nebular age.  

Consequently, we over-plotted on the \citet{2016A&A...588A..25M} tracks a series of CSPN effective temperatures, from 
50 to 300\,kK in steps of 10\,kK, and their corresponding luminosity for the derived CSPN absolute magnitude, in an 
attempt to estimate the CSPN effective temperature \citsee{2022ApJ...935L..35F}. We found the CSPN effective temperature $T_{\rm eff}=130\pm12$\,kK (the error indicates the nebular age and CSPN magnitude uncertainties), well within the 
range of what expected for a CSPN of a Type~I bipolar PN \citep[e.g.][]{1995A&A...293..871C}. In a further and independent attempt to estimate the CSPN effective temperature, we followed the Zanstra method \citep{1931ZA......2....1Z}. From the absolute nebular H$\beta$ flux, the CSPN V magnitude corrected for interstellar extinction and the \ion{He}{ii}$\lambda4686\AA$/H$\beta$ ratio from \citet{1984ApJ...287..341D}, we found a Zanstra 
$T_{\rm eff}(\ion{H}{i})=128\pm39$\;kK and 
$T_{\rm eff}(\ion{He}{ii})=155\pm47$\;kK (an error of 30\% is adopted for the Zanstra temperatures) in excellent agreement with the value estimated from the evolutionary tracks. \citet{2023ApJ...945...11R} find a CSPN effective temperature of 190\,kK from SED fitting. According to our calculations, for the CSPN to be that hot, it would require the nebular age to be less 
than 1100\,yr, and thus either its expansion velocity should be larger than 700{\kms} or its distance less than 330\,pc, which are not supported from our evidence. We believe that this discrepancy is a result of the limited points and the outdated CSPN visual magnitudes from \citet{1988A&A...197..266G} that \citet{2023ApJ...945...11R} uses for the SED fitting and thus, their estimates cannot be reliable. For our derived CSPN effective temperature of 130\,kK, we estimated a CSPN luminosity of $\log \rm L/{\rm L}_{\sun}=2.46\pm0.14$.

From the corresponding isochrone for the cluster's adopted physical parameters, and taking into account the time required for a star to leave the MS and reach the post-Asymptotic Giant Branch (post-AGB) phase, we estimated a CSPN initial mass of 2.33 $\pm$ 0.10~M$_\sun$. The quoted error here reflects the uncertainties of the cluster's physical parameters employed for the isochrone fit. The physical parameters of the PN NGC~2818 and its host cluster NGC~2818A, are summarised in Table~\ref{tab:Tab2}.

\begin{table*}
\caption{PN NGC~2818 and NGC~2818A OC physical properties.}
\centering
\label{tab:Tab2}
\begin{tabular}{lcc}
\hline\hline
         Parameter & PN NGC~2818 & Cluster NGC~2818A \\ 
         \hline \\
         RA (J2000) & 09:16:01.70 & 09:16:10.90 \\
         DEC (J2000) & -36:37:38.75 & -36:38:02.00 \\
         Apparent size (arcsec$^2$) & 140$\times$55 \tablefootmark{(1)} & 450 (diameter) \\
         Physical size (pc) & 2.24($\pm$0.02)$\times$0.88($\pm$0.01) & 3.6 $\pm$ 0.4 (radius) \\
         Morphology & Bipolar & Open Cluster \\
         Radial Velocity ({\kms}) & 23.7, $\sigma$ = 5.5 & 22.0 $\pm$ 1.3  \\
         Distance (kpc) & 3.0 $\pm$ 0.8\tablefootmark{(2)} & 3.30 $\pm$ 0.03 \\
         Reddening $E(B-V)$ & 0.17 $\pm$ 0.08\tablefootmark{(3)} & 0.18 $\pm$ 0.02\tablefootmark{(4)} \\
         Polar expansion velocity ({\kms}) & 91, $\sigma$ = 12 &\\
         Equatorial expansion velocity ({\kms}) & 42, $\sigma$ = 6 &\\
         Chemistry & Type I & [Fe/H]= -0.024 $\pm$ 0.015\tablefootmark{(5)} \\
         Age & $11 \pm 2 $~kyr & 955 $\pm$ 117~Myr  \\
         Cluster Turn-off Mass (M$_\sun$) & & 1.88 \\
         Nebular ionised mass (M$_\sun$) & > 0.47 \tablefootmark{(6)}  & \\
         Nebular electron temperature $T_e[NII]$ (K) & 11500$\pm800$\tablefootmark{(7)}   &  \\
         Nebular electron density $N_e$ (cm$^{-3}$)  &  < 100 \tablefootmark{(8)} &  \\
         Nebular absolute flux $log(H\beta)$ ($mW/m^2$) & -11.28\tablefootmark{(9)} &  \\         
         CSPN T$_{eff}$ (kK) & 130 $\pm$ 12 \tablefootmark{(10)} & \\
         CSPN Zanstra T(HI) (kK) & 128 $\pm$ 39 & \\
         CSPN Zanstra T(HeII) (kK) & 155 $\pm$ 47 & \\
         Estimated CSPN Luminosity 
         logL/L$_\sun$& 2.46 $\pm$ 0.14 \tablefootmark{(11)} & \\  
         Proper motion pmRA (mas/yr)   & -3.71 $\pm$
         0.19  & -4.421 $\pm$ 0.003\tablefootmark{(12)}  \\
         Proper motion pmDec (mas/yr) & 4.94 
         $\pm$ 0.18  &  4.548  $\pm$ 0.003 \tablefootmark{(12)} \\
         CSPN $V$ magnitude & 19.04 $\pm$ 0.10 & \\
         CSPN absolute magnitude $M_V$ & 5.9 $\pm$ 0.3 \tablefootmark{(13)} & \\
         CSPN initial mass (M$_\sun$) & 2.33 $\pm$ 0.10  & \\
         CSPN final mass (M$_\sun$) &  0.58 $\pm$ 0.01  &\\ 
         \hline
  \end{tabular}

    \tablefoot{(1) The mean from H$\alpha$ and [\ion{N}{ii}]$\lambda6584\AA$ imaging from \citet{2024MNRAS.530.3327D}.  
    (2) Refers to the statistical PN distance from \citet{2016MNRAS.455.1459F}.   
    (3) As obtained from \citet{2016MNRAS.455.1459F}.
    (4) Using the \citet{1989ApJ...345..245C} extinction law with Rv=3.1.
    (5) metallicity $Z=0.0144\pm0.0005$ for Z$_\sun=0.0152$.
    (6) Estimated following \citet{2015ApJ...803...99C}.
    (7) From \citet{1984ApJ...287..341D}.
    (8) Average electron density derived from the observed [\ion{S}{ii}]$\lambda6731/6716\AA$ ratio from all pointings assuming $T_e=11500$\,K.
    (9) As obtained from \citet{2023A&A...680A.104B}.
    (10) Estimated from the \citet{2016A&A...588A..25M} evolutionary tracks for $Z=0.01$.
    (11) For $T_{\rm eff}=130$\,kK
    (12) From Gaia DR3 average of cluster members with membership probability $>0.9$ from \citet{2023A&A...675A..68V}.
    (13) As derived for the adopted cluster distance and nebular reddening.
}
\end{table*}

The initial-to-final-mass relation (IFMR), associates the physical characteristics of WDs to these of their progenitor stars and is mainly constructed from cluster WDs \citeg{2018ApJ...866...21C}, since cluster membership allow the determination of their initial masses and properties. A well-established IFMR is essential for depicting the development of vital elements, such as carbon and nitrogen, in entire galaxies. Any additional data point, especially from PNe physically associated with star clusters, offers a significant contribution to the currently, poorly constrained IFMR. In Figure \ref{Fig6}, we over-plotted the initial and final masses of the CSPN of NGC~2818, along with the latest IFMR estimates and semi-empirical `PARSEC' fit \citep{2018ApJ...866...21C}, plus the corresponding data points from the three other PNe currently shown to reside in OCs \citep{2019MNRAS.484.3078F,2019NatAs...3..851F,2022ApJ...935L..35F}. 

   \begin{figure}
\centering
    \subfloat{\includegraphics[width = 3.6in]{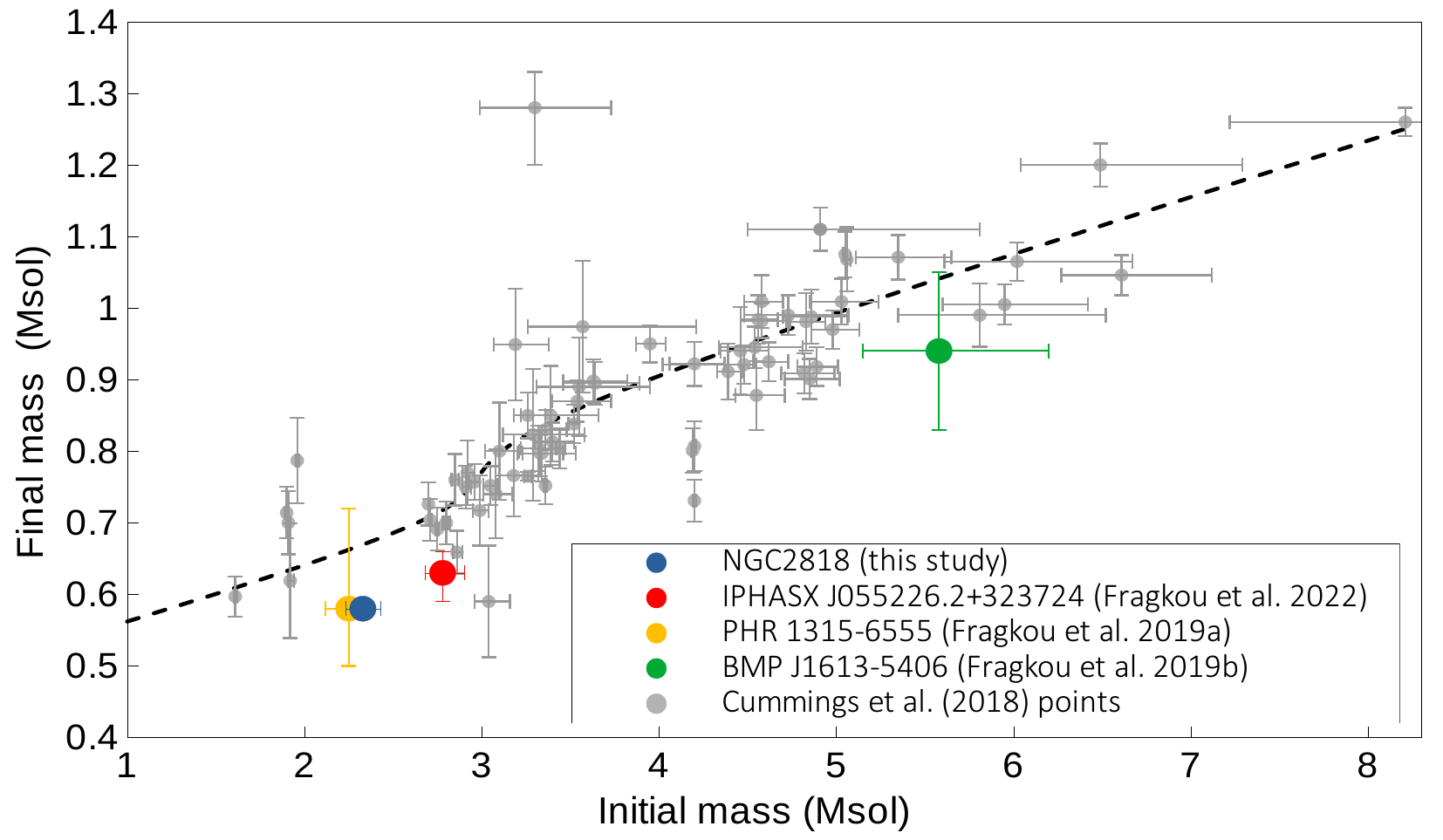}} \\
   \subfloat{\includegraphics[width = 3.6in]{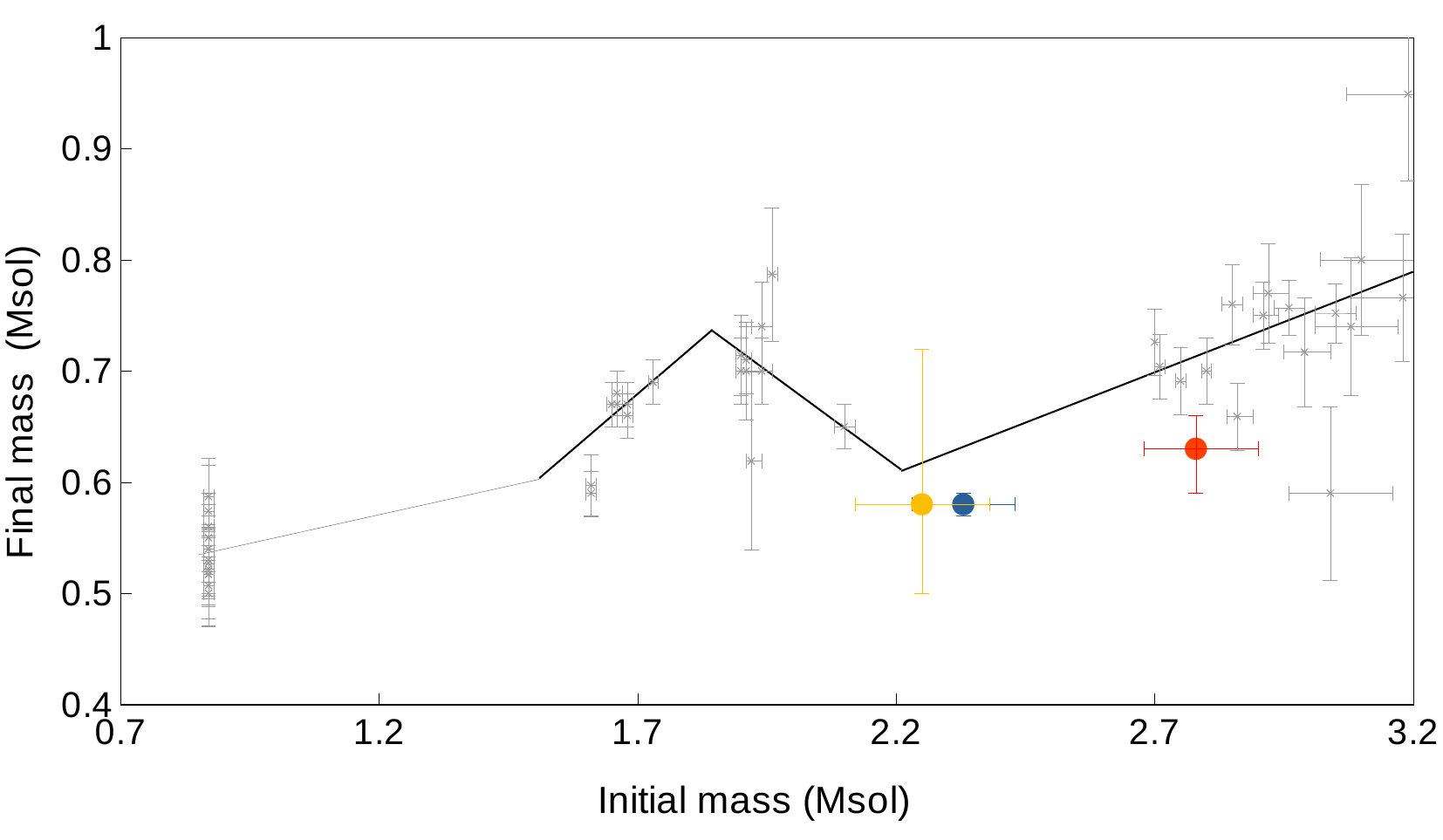}} \\
      \caption{IFMR. The cluster-PN initial and final masses over-plotted along the IFMR data points (in grey) and `PARSEC' trend (dotted line) from \citet{2018ApJ...866...21C} (upper panel) and the \citet{2020NatAs...4.1102M} data points (in grey) and fit (continuous line) depicting the `kink' between 1.5 and 2.2{\msol}(lower panel).} 
         \label{Fig6}
   \end{figure}

Our new data point (in blue) also falls in an important gap of the IFMR and generally agrees with the latest IFMR estimates 
and the data points for the three other PNe in OCs currently known. Our data for the CSPN of PN NGC~2818 falls almost at 
the same position as the CSPN of PN PHR~1315-6555 in OC Andrews-Lindsay 1 (AL1, yellow point), recently reconfirmed to be a true OC-PN association based on HST proper motions estimates of the CSPN \citep{2025AJ....169..199B}. All the previous three OC-PN final masses fall a little lower than the general trend as is the case here too. Moreover, our new data point further confirms and 
delineates the upwards `kink' found by \citet{2020NatAs...4.1102M} that is proposed to be a result of carbon star formation during the AGB phase. Our new data point is located on a sparsely populated area of the IFMR just beyond this kink implying that the kink does not extend to higher initial masses. More data filling the gap between 2.2. and 2.7 initial masses will help with providing context for the underlying mechanisms that are responsible for this peculiarity of the IFMR. 

The PN N/O abundance of $0.99\pm0.03$ determined by \citet{2023A&A...680A.104B}, does not agree with the predicted yields for a progenitor star of 2.33{\msol} \citep{2016ApJ...825...26K}. This is something that has to be 
further explored in future studies. It is possibly associated with the mechanisms that produce the kink depicted in the IFMR and perhaps the likely binary nature of at least the CSPN of NGC\,2818 and likely several of the other OC-PN too.

\section{Conclusions}
\label{sec:conc}
Creating a link between the properties of a PN with those of its progenitor star is of vital importance. PNe that are 
physical members of star clusters can provide us with more robust estimates of key such properties, offering additional insights into stellar evolution. However, PNe in Galactic OCs
are very rare and any new discovery is significant. We have demonstrated, with robust arguments, that the PN NGC\,2818 is likely to be a physical member of the Galactic OC NGC\,2818A, an association long debated in previous studies but further secured here. All crucial parameters for establishing an association of a cluster with a PN agree to within the often reduced errors from the new data presented in this study. These include agreement of distance of PN and cluster, location of PN within the cluster tidal radius, plausible cluster physical size, reddening agreement, and most importantly tight agreement of radial velocities given the 1\,{\kms} OC velocity dispersions \citep[see][]{2019NatAs...3..851F, 2022ApJ...935L..35F}. Taken together these present compelling evidence of true association. Future observations, including deep spectra of the central star, would provide the definitive confirmation needed to strengthen this particular argument. However, such observations are beyond most existing facilities' capabilities due to the central star's faintness.

We derived the physical parameters of the cluster, the PN and its CSPN along with these of its progenitor star. Our estimates agree with these from previous studies and with the latest IFMR estimates. Our new data point populates an important gap of the IFMR confirming and further limits the scope of the kink found in the IFMR by \citet{2020NatAs...4.1102M} by showing that it does not exceed into higher initial masses. The PN chemical yields have to be further explored, since they do not agree with current models and this is an important consideration. 

So far all identified OC-PNe are evolved bipolars, nitrogen enriched and have relatively high-mass progenitors (see Table 
\ref{tab:Tab3}). This is as expected given the OC age range and so the associated and necessitated high TO masses. Four PN in this OC-PN class have now been found, with one OC-PN having an intermediate progenitor mass that is close to the limit of core-collapse supernova formation, and another OC-PN being the oldest 
confirmed PN ever found. Interestingly, all their final masses fall below the general IFMR trends from cluster WDs (but just about compatible agree within the errors) and can be effectively joined by a straight line. We have two further cases under study hoping to add in this rare class soon to further explore these issues. 

\begin{table*}
\caption{Properties of the identified OC-PNe.}
\centering
\label{tab:Tab3}
\begin{tabular}{lcccccccc}
\hline\hline
          &  \multicolumn{2}{c|}{PN in AL1\tablefootmark{(1)}} & \multicolumn{2}{c|}{PN in NGC~6067\tablefootmark{(2)}} & \multicolumn{2}{c|}{PN in M37\tablefootmark{(3)}} & \multicolumn{2}{c|} {PN in NGC~2818A} \\           
\hline
         Parameter & PN & Cluster & PN & Cluster & PN & Cluster & PN & Cluster \\
         \hline \\
         RA (J2000) & 13:15:19 & 13:15:16 & 16:13:02 & 16:13:11 & 05:52:26.2 & 05:52:18 & 09:16:02 & 09:16:11 \\
         DEC (J2000) & $-65$:55:01 & $-65$:55:16 & $-54$:06:32 & $-54$:13:06 & 32:37:25 & 32:33:12 & $-36$:37:39 & $-36$:38:02 \\
         Morphology & Bipolar & OC & Bipolar & OC & Bipolar & OC & Bipolar & OC \\
         Physical radius (pc) & 0.42 & & 1.27 & 8.10 & 1.60 & 6.72 & 2.24$\times$0.88\tablefootmark{(4)} & 3.6 \\
         Chemistry & Type I & 0.006\tablefootmark{(5)} & Type I & 0.024\tablefootmark{(5)} & Type I & 0.016\tablefootmark{(5)} & Type I & 0.014\tablefootmark{(5)}  \\
         Age & 11~kyr & 0.66~Gyr & 31~kyr & 90~Myr & 78~kyr & 470~Myr & 11~kyr & 955~Myr  \\
         Ionised mass (\msol) & 0.23 & & 0.56 & & 0.32 && > 0.47 & \\
        $N_e$ (cm$^{-3}$)  &  160 & & < 5 & & 5 & & <100  &  \\  
         CSPN $T_{\rm eff}$ (kK) & 113 && 125-190 && 100 && 130 & \\
         CSPN $M_{\rm init}$ (\msol) & 2.25$\pm 0.13$ && 5.58$_{-0.43}^{+0.62}$ &&  2.78$_{-0.10}^{+0.12}$ &&  2.33 $\pm$ 0.10  & \\
         CSPN $M_{\rm fin}$ (\msol) & 0.58$_{-0.08}^{+0.14}$ && 0.94$\pm 0.11$ && 0.63$_{-0.04}^{+0.03}$ && 0.58 $\pm$ 0.01  &\\ 
         \hline
  \end{tabular}
    \tablefoot{(1) From \citet{2011MNRAS.413.1835P} and \citet{2019MNRAS.484.3078F}. 
    (2) From \citet{2019NatAs...3..851F}.
    (3) From \citet{2022ApJ...935L..35F}.\\
    (4) major $\times$ minor axis.
    (5) metallicity $Z$ for $Z_\sun=0.0152$.
}

\end{table*}

\section{Data availability}
The Gemini spectral data are only available in electronic form at the CDS via anonymous ftp to cdsarc.u-strasbg.fr (130.79.128.5) or via http://cdsweb.u-strasbg.fr/cgi-bin/qcat?J/A+A/.

\begin{acknowledgements}
We thank the anonymous referee for pointing out important aspects improving the manuscript. VF thanks Fundação Coordenação de Aperfeiçoamento de Pessoal de Nível Superior (CAPES) for granting the postdoctoral research fellowship Programa de Desenvolvimento da Pós-Graduação (PDPG)-Pós Doutorado Estratégico, Edital nº16/2022. This study was financed in part by the Coordenação de Aperfeiçoamento de Pessoal de Nível Superior - Brasil (CAPES) – Finance Code  001 (88887.838401/2023-00). VF was supported by a UNAM postdoctoral fellowship at the beginning of this study. VF and RV were supported by UNAM-PAPIIT grant IN106720. QAP thanks the Hong Kong Research Grants Council for GRF research support under grants 17326116 and 17300417. DRG acknowledges FAPERJ (E-26/200.527/2023) and CNPq (315307/2023-4) grants. LLN thanks Funda\c{c}\~ao de Amparo \`a Pesquisa do Estado do Rio de Janeiro (FAPERJ) for granting the postdoctoral research fellowship 
E-40/2021(280692). 

Based on observations obtained at the international Gemini Observatory, a program (Program ID: GS-2023B-FT212) of NSF NOIRLab (acquired through the Gemini Observatory Archive), at NSF NOIRLab and processed using the Gemini IRAF package and DRAGONS (Data Reduction for Astronomy from Gemini Observatory North and South), which is managed by the Association of Universities for Research in Astronomy (AURA) under a cooperative agreement with the U.S. National Science Foundation on behalf of the Gemini Observatory partnership: the U.S. National Science Foundation (United States), National Research Council (Canada), Agencia Nacional de Investigaci\'{o}n y Desarrollo (Chile), Ministerio de Ciencia, Tecnolog\'{i}a e Innovaci\'{o}n (Argentina), Minist\'{e}rio da Ci\^{e}ncia, Tecnologia, Inova\c{c}\~{o}es e Comunica\c{c}\~{o}es (Brazil), and Korea Astronomy and Space Science Institute (Republic of Korea).

This work made use of the University of 
Hong Kong/Australian Astronomical Observatory/Strasbourg Observatory H-alpha Planetary Nebula (HASH PN) database, hosted by the Laboratory for Space Research at the University of Hong Kong; This research has made use of the SIMBAD database and the VizieR catalogue access tool, CDS, Strasbourg, France (doi: 10.26093/cds/vizier). The original description of the VizieR service was published in Ochsenbein, Bauer \& Marcout (2000).

This work has made use of data from the European Space Agency (ESA) mission Gaia (https://www.cosmos.esa.int/gaia), processed by the Gaia Data Processing and Analysis Consortium (DPAC, 
https://www.cosmos.esa.int/web/gaia/dpac/consortium).

We thank Dr. Stavros Akras for his valuable input in the analysis of the data.

\end{acknowledgements}

\bibliographystyle{aa} 
\bibliography{ngc2818}

\end{document}